\documentclass[aps,pre,amsmath,superscriptaddress,12pt]{revtex4}
\usepackage{graphicx}

\def\eq#1{Eq.~(\ref{#1})}
\def\fig#1{Fig.~\ref{#1}}

\begin{document}

\title{Elastic interactions of active cells with soft materials}
\author{I. B. Bischofs}
\affiliation{Max Planck Institute of Colloids and Interfaces, 14424 Potsdam, Germany}
\author{S. A. Safran}
\affiliation{Weizmann Institute of Science, Rehovot 76100, Israel}
\author{U. S. Schwarz}
\email[To whom correspondence should be addressed. Email: ]{Ulrich.Schwarz@mpikg-golm.mpg.de}
\affiliation{Max Planck Institute of Colloids and Interfaces, 14424 Potsdam, Germany}
\affiliation{Institute of Theoretical Physics, University of Leipzig, 04103 Leipzig, Germany}
            
\begin{abstract}
  Anchorage-dependent cells collect information on the mechanical
  properties of the environment through their contractile machineries
  and use this information to position and orient themselves.  Since
  the probing process is anisotropic, cellular force patterns during
  active mechanosensing can be modelled as anisotropic force
  contraction dipoles. Their build-up depends on the mechanical
  properties of the environment, including elastic rigidity and
  prestrain. In a finite sized sample, it also depends on sample
  geometry and boundary conditions through image strain fields.  We
  discuss the interactions of active cells with an elastic environment
  and compare it to the case of physical force dipoles.  Despite
  marked differences, both cases can be described in the same
  theoretical framework.  We exactly solve the elastic equations for
  anisotropic force contraction dipoles in different geometries (full
  space, halfspace and sphere) and with different boundary
  conditions. These results are then used to predict optimal position
  and orientation of mechanosensing cells in soft material.
\end{abstract}

\maketitle

\newpage

\section{Introduction}

Anchorage-dependent cells like fibroblasts in connective tissue show a
remarkable degree of mechanical activity. The first quantitative
measurements of cellular traction were performed with the elastic
substrate method in the early 1980s by Harris and coworkers, who found
that cells exert much larger forces than previously thought
\cite{c:harr80,c:harr81}.  During recent years, the elastic substrate
method has been improved considerably
\cite{c:demb96,c:demb99}.  In particular, a new variant involving
micro-patterning has been developed, that allows to resolve individual
forces exerted at single focal adhesions
\cite{uss:bala01,uss:schw02b}. \textit{Focal adhesions} are mature 
adhesion contacts based on transmembrane proteins from the integrin
family.  Since they connect the extracellular matrix and the actin
cytoskeleton, they can transmit internal forces to the environment and
external forces to the cell. Using micro-patterned elastic substrates,
it was found that fibroblasts typically exert forces of 10 nN at
mature focal adhesions \cite{uss:bala01,uss:schw02b}. Using a bed of
flexible microneedles, similar values were found for smooth muscle
cells \cite{c:tan03}. Since adherent cells can have up to hundreds of
focal adhesions, the overall force exerted by the cell can amount to
$\mu$N.  The forces exerted by cells on their environment result from
non-equilibrium processes inside the cell and are generated by myosin II
molecular motors interacting with the actin cytoskeleton. Since
typical forces produced by molecular motors are in the pN-range
\cite{b:howa01}, there must be up to $10^6$ myosin II molecular motors
contributing to overall cell traction.

When Harris and coworkers first discovered these large forces, they
concluded that they are required for the physiological function of the
specific cell type under consideration. For example, fibroblasts are
believed to maintain the integrity of connective tissue by
mechanically pulling on the collagen fibers. Moreover, they are an
integral part of the wound contraction process. Harris and coworkers
also noticed that cells react to mechanical changes in their
environment caused by traction of other cells.  Since cells are known
to align along topographic features in their environment
(\textit{contact guidance}), they suggested that cells react to
traction-induced reorganization of collagen fibers. This mechanism
amounts to a mechanical interaction of cells and has been addressed
theoretically in coupled transport equations for fiber and cell
degrees of freedom \cite{c:oste83,c:baro97}.

During recent years, the sophisticated use of elastic substrates has
shown that cells also react to purely elastic features in their
environment, including rigidity, rigidity gradients and prestrain
\cite{c:pelh97,c:lo00,c:wong03}. It is now generally accepted that
these effects are related to the special properties of focal adhesions
\cite{c:geig02}. In particular, it has been shown
that application of external force leads to growth of
focal adhesions and therefore to strong signaling activity
\cite{c:wang93,c:choq97,uss:rive01}.  The same aggregation has been
found for mature focal adhesions under internally generated force
\cite{uss:bala01,uss:schw02b,c:tan03}, suggesting that focal adhesions
act as mechanosensors that convert force into biochemistry and vice
versa. Therefore the mechanical activity of cells is not only related to
the physiological function of their cell type, but is also a general
way to collect information about the mechanical properties of the
environment (\textit{active mechanosensing}). There is strong evidence
that this mechanism is involved in many important physiological
situations, including tissue maintenance, wound healing, angiogenesis,
development and metastasis \cite{c:chic98,c:galb98,c:huan99}.

\begin{figure}
\begin{center}
\includegraphics[width=0.8\columnwidth]{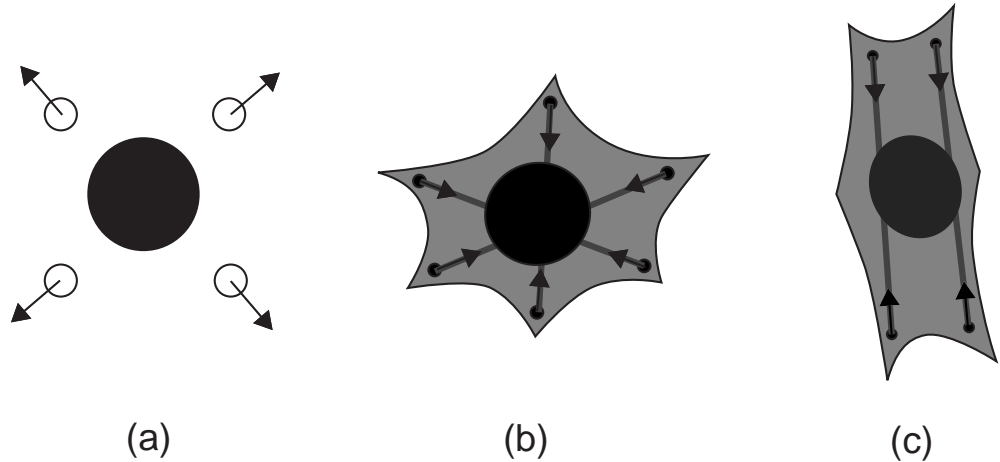}
\caption{\label{cartoon} Schematic representation of physical and 
cellular force dipoles. (a) Physical case: an intercalated defect
deforms the simple cubic host lattice, thus acting as an isotropic
force expansion dipole. (b) Cellular case: anchorage-dependent cells
probe the mechanical properties of the soft environment through their
contractile machinery. Actin stress fibers (lines) are contracted by
myosin II molecular motors and are connected to the environment
through focal adhesions (dots). Even if cell morphology is round or
stellate, different stress fibers probe different directions of space
and compete with each other for stabilization of the corresponding
focal adhesions. Therefore the probing process can be modeled as
anisotropic force contraction dipole. (c) Cell morphology becomes
elongated in response to anisotropic external stimuli, during
locomotion or spontaneously during times of strong mechanical
activity.  Then most stress fibers run in parallel and the whole cell
appears as an anisotropic force contraction dipole.}
\end{center}
\end{figure}

The dynamics of focal adhesions is a subject of much current research
\cite{c:geig01a}. Anchorage-dependent cells constantly assemble and
disassemble focal adhesions, thereby probing the mechanical properties
of their environment. Initial focal adhesions (\textit{focal
complexes}) are local processes based on integrin clustering. If
initial clustering is stabilized by the properties of the
extracellular environment, focal complexes can mature into focal
adhesions. In this case, they connect to the actin cytoskeleton and a
contractile force pattern builds up, that is actively generated by
myosin II molecular motors interacting with the actin
cytoskeleton. The minimal configuration of this machinery is a set of
two focal adhesions connected by one bundle of actin filaments
(\textit{stress fiber}), that leads to a pinch-like force pattern. In
condensed matter physics, such an object is known as an
\textit{anisotropic force contraction dipole} \cite{e:siem68}.
The concept of force dipoles has been applied before mainly for the
description of point defects in traditional condensed matter systems,
including hydrogen in metal (e.g.\ platinum) \cite{e:wagn74}, atoms
adsorbed onto crystal faces (e.g.\ argon on gold)
\cite{e:lau77}, or intercalation compounds (e.g.\ lithium in graphite)
\cite{e:safr79}. The concept of force dipoles has also been used to model 
active biological particles in a fluid environment, e.g.\ ion pumps
\cite{c:rama00} and rotary motors \cite{c:lenz03a} embedded in fluid
membranes, or self-propelled particles like swimming bacteria
\cite{c:simh02}. Recently, we have suggested to use the concept of
force dipoles to model the mechanical activity of cells
\cite{uss:schw02a}. Cells in an isotropic environment often show
isotropic (that is round or stellate) morphologies. However, since the
focal adhesion dynamics is local, even in this case there is an
anisotropic probing process, that can be modeled by anisotropic force
contraction dipoles. As we will argue below, only an anisotropic
probing process can react to anisotropies in the environment.  The
anisotropy of focal adhesion dynamics becomes apparent when stress
fibers start to orient in one preferential direction, either
spontaneously during a period of large mechanical activity, or as a
response to some external anisotropy, or during cell locomotion.  In
this case, cellular dipoles have been measured to be of the order of
$P \approx - 10^{-11} J$ (this corresponds to two forces of 200 nN
each, separated by a distance of 60 $\mu$m)
\cite{uss:schw02b,c:butl02}. In \fig{cartoon} we show schematic
representations of the physical and cellular cases discussed here.

In order to sense the mechanical properties of their environment,
cells can make use of the fact that these properties modulate the
build-up of their own force patterns. In this paper, we focus on the
role of stress and strain in the extracellular material for cellular
decision making in regard to positioning and orienting in a soft
environment.  In order to calculate how stress and strain are
propagated in the environment, the extracellular material is modeled
using isotropic linear elasticity. This is certainly true for
synthetic elastic substrates (usually made from polydimethylsiloxane
or polyacrylamide). The typical physiological environment for
anchorage-dependent cells are hydrogels, whose mechanical properties
are more difficult to model, in particular due to their viscoelastic
and non-linear behaviour. Yet our calculations will show that our
model has large predictive power also for this case, possibly because
elastic deformations of hydrogels become encoded in plastic changes
that later can be detected by active mechanosensing in a similar way
as persistent elastic deformations.  Given the assumption of isotropic
linear elasticity, we can calculate how stress and strain follows from
the force dipoles by solving the elastic equations for the geometry
and boundary conditions of interest. 

The most critical part of our modeling is the way in which physical or
cellular force dipoles react to stress and strain in their
environment.  This subject has been treated extensively for the case
of atomic defects in traditional condensed matter systems
\cite{e:wagn74,e:lau77,e:safr79}.  Here defects are usually modeled as
isotropic force expansion dipoles. The equilibrium configuration
follows by minimizing the sum of the elastic energy of the strained
medium and the direct interaction energy between force dipole and
elastic environment. The first term represents a restoring force and
raises the energy (i.e.\ its sign is always positive), while the
second term is a driving force that reduces the total energy (i.e.\
its contribution will always be negative). The equilibrium
configuration will correspond to the minimum of the total energy as a
function of position and orientation of the force dipoles, which
results in an effective, so-called
\textit{elastic interaction} between the force dipole and other
dipoles, sample boundaries or external strain fields. One central
result of these studies is that the direct interaction between
isotropic force dipoles in an isotropic elastic material vanishes
\cite{e:siem68} and that they interact through a boundary-induced 
(\textit{image}) interaction that varies on the length scale of the
sample size (leading to \textit{macroscopic modes})
\cite{e:wagn74}. For anisotropic force dipoles, the direct elastic
interaction does not vanish. Recently, we have predicted that the
competition between direct and image interactions should lead to
hierarchical structure formation, with the direct interaction leading
to structure formation on a length scale set by the elastic constants
and similar to that of electric quadrupoles \cite{uss:schw02a}. We
suggested that such a behaviour should be expected for artificial or
inert cells, that is for physical particles with a static force
contraction dipole, but without any internal dynamic or regulatory
response.  

In contrast to this physical case, the effective behavior of active
cells usually follows from dynamic and tightly regulated
non-equilibrium processes inside the cell.  More recently, we have
shown that despite this severe complication, it is still possible to
describe the active response of mechanosensing cells in an elastic
material in the same framework as the physical case
\cite{uss:bisc03a}. In detail, motivated mainly by recent
experiments with elastic substrates \cite{c:pelh97,c:lo00,c:wong03},
we have suggested that effective cellular behavior can be described as
simple preference for large effective stiffness in the environment
(including both rigidity and tensile prestrain). Moreover we have
shown that this principle is equivalent to minimization of the energy
which the cells have to invest into straining the environment in order
to build-up the force dipole used for probing the mechanical
properties of the environment. One likely explanation for the observed
active behavior of cells is that the build-up of force at focal
adhesions is more efficient in a stiff environment. Since this
approach allows to use the same framework as in the physical case, we
were able to derive elastic interaction laws between cells and their
elastic environment which are in good agreement with experimental
observations for fibroblasts both on elastic substrates and in
hydrogels. In particular, the direct elastic interaction between cells
has been predicted to be similar to that of electric dipoles, leading
to strings of cells \cite{uss:bisc03a}.

In this paper, we present a unifying formalism for theoretical models
for elastic interactions for both physical force dipoles and active
cells. In particular, we consider interactions with external strain
fields, sample boundaries or other physical force
dipoles/cells. Although there are marked conceptual differences
between the physical and cellular cases, they both require to solve
the elastic boundary value problem to predict the resulting structure
formation. Since cells are modeled as anisotropic force dipoles, these
calculations are in general more involved than similar calculations
for isotropic force dipoles. Moreover, in contrast to earlier
calculations for the physical case, we are interested not only in the
effect of free, but also of clamped boundaries, which are known to
induce mechanical activity of cells \cite{c:grin00}. Our paper is
divided into two parts. In the first part, we discuss the details of
our modeling, in particular the difference between physical and
cellular force dipoles. In the second part, we apply our model to
several cases of interest.  Here we present exact solutions of the
elastic equations for anisotropic force dipoles in full space,
halfspace, and sphere, and apply them both to physical and cellular
force dipoles. For example, we show that cells are attracted and
repeled by clamped and free sample boundaries, respectively. In the
case of physical force dipoles, this behaviour is inverted.  Our
predictions for cells explain many experimental findings reported in
the literature, can be used for rational design of tissue equivalents,
and show that physical concepts can provide new and important insight
into cell biology, provided that they are applied with adequate
modifications.

\section{Modeling}

\subsection{Force multipoles}

In the following, we model a mechanically active cell as a localized
force distribution in an elastic medium. In order to
describe its mechanical action, we use the concept of a force
multipolar expansion, which has been applied before for the
description of point defects in condensed matter systems
\cite{e:wagn74,e:lau77,e:safr79}. Consider a force distribution
localized around the origin.  Then the force multipoles are defined as
\cite{e:siem68}
\begin{equation} \label{eq:force_multipoles}
  P_{i_1 \dots i_n i} = \int s_{i_1} \dots s_{i_n} f_i(\vec s)\ d^3s
\end{equation}
where $f_i$ is the force density and $d^3 s$ denotes a volume
integral. The first order term is the vector of overall force, $P_i$,
and the second order term is the force dipole, $P_{ij}$, a tensor of
rank two. For both the cellular and physical situation we are
interested in, we can assume local forces. For point-like defects, one
can moreover assume that the overall force vanishes, because due to
Newton's Third Law, the forces exerted by the defect on the elastic
medium and by the elastic medium on the defect have to balance each
other (the same argument applies to point defects in a fluid medium
\cite{c:rama00,c:lenz03a,c:simh02}). For cells, the situation
is more complicated, because they are at the same time in contact with
the elastic matrix and an aqueous medium, thus unbalanced forces might
appear in the elastic matrix, which are balanced by viscous forces in
the aqueous medium. However, viscous processes in the fluid medium
decay very rapidly on the timescale of cell movement. Therefore
unbalanced forces might occur for short periods of time, e.g.\ during
back retraction of locomoting cell, but during most of the time, cells
can be expected to be in mechanical equilibrium, as suggested by
experiments measuring force patterns of both stationary and locomoting
cells on elastic substrates \cite{c:demb99,uss:schw02b}. Our model for
cellular force patterns can be interpreted as one stress fiber
connecting two focal adhesions. Obviously this minimal system
obeys mechanical equilibrium.  Then overall force vanishes and the
force dipole is the first relevant term in the multipolar expansion
\eq{eq:force_multipoles}.  

Force dipoles are classified according to their symmetry properties
into isotropic dipoles (centers of contraction or dilation),
anisotropic dipoles without moment and anisotropic force dipoles with
moment \cite{b:love44}.  Force dilatation and force contraction
dipoles have only positive and only negative eigenvalues,
respectively.  For example, in three dimensions three pairs of
juxtaposed forces, one for each coordinate direction, form an
isotropic force dipole, where $P_{ij} = P \delta_{ij}$. Such a force
dipole describes a spherical inclusion in a simple cubic lattice, see
\fig{cartoon}a \cite{e:wagn74}. Applied to two dimensions, it
describes atomic defects adsorbed onto a substrate
\cite{e:lau77}. An anisotropic force dipole without moment is a
non-diagonal, but symmetric tensor.  For example, for a couple of
juxtraposed forces with a dipole strength $P$ and an orientation in
direction $\vec l$, we can write the force dipole tensor as $P_{ij} =
P \hat{l}_i \hat{l}_j$. Such dipoles are used below to describe the
probing force patterns of cells, see \fig{cartoon}b and c
\cite{uss:schw02a}. An anisotropic force dipole without moment
oriented in the z-direction reads $P_{ij} = P
\delta_{iz} \delta_{jz}$ and describes for example an atomic defect
intercalated in graphite \cite{e:safr79}. Finally, an anisotropic
force dipole with an angular moment describes a set of two opposing
forces $\vec{F}$ separated by a distance $\vec{l}$ oriented
arbitrarily with respect to $\vec{F}$, which leads to $P_{ij}\neq
P_{ji}$. In this paper, we only consider force dipoles without
such moments.

\subsection{Interaction between physical dipoles and an elastic medium}

The elastic medium surrounding a particle can mediate an
\textit{elastic} interaction with other particles, sample boundaries 
or external strain fields. It is important to note that this effect
requires a \emph{direct} interaction of the particle with its elastic
environment.  In traditional condensed matter systems, the direct
interaction is usually a quantum effect (e.g.\ Born repulsion for
defects intercalated into a crystal lattice or van der Waals
attraction for defects adsorbed onto a crystal lattice).  The
interaction of a single particle localized at $\vec{r}$ with the
elastic medium can be described by an interaction potential $V_d(\vec
{r},\vec{u})$, which not only depends on position $\vec{r}$, but which
also is a functional of the displacement field $\vec{u}(\vec{r'})$ of
the elastic medium.  For a fixed particle position $\vec{r}$, we can
expand the interaction potential with respect to the displacement
field:
\begin{equation} \label{potential}
V_d(\vec{r}, \vec{u}) \approx - \int f_i(\vec{r}+\vec{s}) u_i(\vec{r}+\vec{s})\ d^3 s\ ,
\end{equation}
where $f_i = - \delta V_d / \delta u_i|_{u_i=0}$ is the force density
exerted by the defect onto the elastic medium in its undeformed
reference state. Here and in the following, summation over repeated
indices is always implied. The expansion can be terminated after the
linear term because we assume small deformations, or, equivalently,
small forces. This linearized interaction potential is widely used in
the literature on elastic defects in traditional condensed matter
materials \cite{e:wagn74,e:lau77,e:safr79}. For later use, we also
note that \eq{potential} can be rewritten in terms of the force
multipoles defined in \eq{eq:force_multipoles}, if one makes the
assumption that the interaction of the defect with the medium is
short-ranged. Then
\begin{equation} \label{eq:vd_effective}
V_d(\vec{r},\vec u) \approx - \sum_{n=0}^{\infty} \frac{1}{n !} 
P_{i_1 \dots i_n i} u_{i,i_1 \dots i_n}(\vec{r})\ ,
\end{equation}
where indices after the comma denote derivatives of the displacement
field with respect to position ($u_{,i} = \partial u / \partial r_i$).
In this way, all the details of the direct interaction between medium
and defect are subsumed in the defect force pattern and one can study
elastic effects in different materials within a common theoretical
framework, as long as the two assumptions of small and localized
forces are valid.

The displacements of the elastic medium are controlled by a
competition between the direct interaction between defect and medium
and the elastic strain energy of the medium under the constraints of
adequate boundary conditions. The strain energy is \cite{b:land70}
\begin{equation} 
V_e = \frac{1}{2} \int d^3r\ C_{ijkl} u_{ij}(\vec{r}) u_{kl}(\vec{r})
\label{elasticWork}
\end{equation}
where $u_{ij}(\vec{r})$ is the strain tensor and $C_{ijkl}$ the elastic
constant tensor of the medium.  Consider now the general case of an elastic
medium subject to loading with defects with an overall volume force
density $\vec f(\{\vec{r}^{\alpha}\}, \vec{r}) = \sum_{\alpha} \vec
f^{\alpha}(\vec{r})$, where $\alpha$ numbers the different
defects. Then the total energy of the system is
\begin{eqnarray}
V_t =  \frac{1}{2} \int d^3r\ C_{ijkl} u_{ij}(\vec{r})
u_{kl}(\vec{r}) - \int d^3r\ f_{i}(\{\vec{r}^{\alpha}\},\vec
r) u_{i} (\vec{r}) - \oint dS\ f^s_{i} (\vec{r}) u_{i} (\vec{r})\ ,
\label{Lagrangian}
\end{eqnarray}
where the first term is the strain energy $V_e$ and the second term
the direct interaction $V_d = \sum_{\alpha}
V_d(\vec{r}^{\alpha})$. The surface force density $\vec f^s$ in the
third term acts as a Lagrange multiplier that enforces the boundary
conditions at the sample surface $S$.  For a fixed defect
configuration, the displacements $\vec u (\vec r)$ are determined from
$\delta V_t / \delta \vec u = 0$, which defines mechanical
equilibrium:
\begin{equation}
C_{ijkl} u_{kl,j}(\vec{r})=-f_i(\{\vec{r}^{\alpha}\}, \vec{r}) \qquad \text{$\vec{r}$ in $V$}, 
\label{equilibrium1}
\end{equation}
and the boundary condition at the surface of the elastic material:
\begin{equation}
C_{ijkl}u_{kl}(\vec{r})n_j(\vec{r}) = f_i^s(\vec{r}) \qquad \text{$\vec  r$ on $S$}, 
\label{equilibrium2}
\end{equation}
where $\vec n$ is the outward directed surface normal of the surface
element dS. By combining Eq.~(\ref{equilibrium1}) and
Eq.~(\ref{elasticWork}), one finds $V_e = \frac{1}{2} \int d^3r\ f_i u_i = -
\frac{1}{2} V_d$.  Therefore the overall energy $V_t = V_d + V_e 
= \frac{1}{2} V_d = - V_e$ and the overall energy can be
written as function of the defect configuration only. In this way, the
\textit{direct} interactions of the particles with the medium can be
rigorously transformed into an \textit{indirect} interaction between
defects. This also allows the calculation of the interaction of a
single defect with a boundary induced strain field or an external strain
field applied at the boundary.  The groundstate configuration of
elastically interacting defects is obtained by minimizing total
energy $V_t$.

\subsection{Interaction between active cells and an elastic medium}

The forces exerted by mechanically active cells on the environment
are mainly due to actomyosin contractility.  Thus, in contrast to the
interaction of physical force dipoles with the elastic medium, where
the force can be derived from conventional interaction potentials,
cellular forces are continuously and actively generated by the cell
and involve non-equilibrium processes, that are tightly regulated by
biochemical events inside the cell.  Hence, the interactions of cells
with an elastic environment are more complicated than for physical
defects and there is little a priori reason why they should be
described by Eq.~(\ref{potential}). Motivated by recent experiments
with elastic substrates \cite{c:pelh97,c:lo00,c:wong03}, we have
argued before that despite these complications, a similar description
as for the physical case can be employed for the cellular one
\cite{uss:bisc03a}. We asked which kind of information a cell can
extract from its elastic environment using its contractile machinery
and suggested that an appropriate scalar quantity to characterize the
environment is the work the cell has to perform in order to build up a
certain level of force against the elastic environment.  Experimental
observations suggest that active cell behaviour amounts to a simple
preference for large effective stiffness, which corresponds to a
minimization of this energy. As a simple analogue, consider a linear
spring. In order to build up a certain force $F$, the energy $W = K
x^2 / 2 = F^2 / 2 K$ has to be invested into the spring, where $x$ is
displacement and $F = K x$ is force at equilibrium. If there is a
choice of different springs with different spring constants $K$, the
smallest amount of energy $W$ to build up $F$ has to be invested into
the spring with the largest value for $K$. In the case of cells, the
different springs correspond to different directions as probed by
different stress fibers, and on the long run, the cell will orient in
that direction that corresponds to the largest value of $K$, possibly
because in this direction, the build-up of force is most
efficient. The example of the linear spring can also be used to
illustrate the main difference to the physical case, when the final
configuration is determined by the overall energy $V_t = K x^2 / 2 - F x
= - F^2 / 2 K = - W$. Thus in contrast to the case of cellular force
dipoles, for physical dipoles minimal values of stiffness $K$ are most
favorable.

We now explain our reasoning in more detail for the case of cells in a
three-dimensional environment described by continuum elasticity
theory. In order to identify a suitable analogue to the spring
constant $K$, we introduce the concept of \textit{local effective
stiffness} of the elastic environment.  We define this quantity to be
the work $W$ required to build up a unit force in the elastic
medium. The deformation work $W$ required to build up an arbitrary
force distribution $\vec f(\vec{r})$ is given by:
\begin{equation}
W= \int d^3r\ \int_0^{u_{ij}^{\vec f}}\ C_{ijkl}u_{kl}(\vec r) du_{ij}(\vec{r}),
\label{work1}
\end{equation}
which in the absence of external prestrain is equivalent to the energy
stored in the elastic medium given in Eq.~(\ref{elasticWork}). Then
integration by parts gives
\begin{equation}
W = -\frac{1}{2}\int d^3r\ u_i(\vec{r}) C_{ijkl} u_{kl,j}(\vec{r})
+ \frac{1}{2} \oint dS\ n_j C_{ijkl} u_{kl}(\vec{r}) u_{i}(\vec r)\ .
\end{equation}
Applying the mechanical equilibrium conditions of the
elastic medium, Eqs.~(\ref{equilibrium1},\ref{equilibrium2}), yields
\begin{equation}
W=\frac{1}{2}\int d^3r\ u_i(\vec{r}) f_i(\vec
r)+\frac{1}{2}\oint dS\ u_{i}(\vec{r}) f_i^s(\vec{r})\ .
\label{work1b}
\end{equation}
In an infinite medium the boundary condition at the surface yields a vanishing
surface integral. Hence for a force distribution centered around $\vec{r}$, the
volume integral can be turned into a local expression by using the 
definitions of Eq.~(\ref{eq:force_multipoles}):
\begin{equation}
 W^{\infty}=\frac{1}{2}\int f_i(\vec{r} + \vec s)\ u_i(\vec{r} + \vec s)\ d^3 s
= \frac{1}{2}\sum_{n=0}^{\infty} \frac{1}{n !} P_{i_1 \dots i_n i} u_{i,i_1
  \dots i_n}(\vec{r}).
\label{referenceEnergy}
\end{equation}
In particular, for a force monopole and a force dipole one finds
$W^{\infty}=\frac{1}{2} P_i u_i^{\infty}(\vec{r})$ and
$W^{\infty}=\frac{1}{2} P_{ij}u^{\infty}_{ij}(\vec{r})$, respectively,
where $\vec u^{\infty}$ and $u_{ij}^{\infty}$ are the displacement and
strain tensor fields caused by the respective force multipole in an
infinite homogeneous medium.  $W^{\infty}$ relates the effective
stiffness encountered by a cell to the elastic constants.  Since
strain scales inversely with elastic constants, $W^{\infty}$ decreases
if the elastic constants increase. For an elastically anisotropic
medium, $W^{\infty}$ varies with the direction of force application,
which provides an orientational clue for cell orientation. As we will
see below, tensile prestrain or boundary-induced tensile image strain
also leads to an increased effective stiffness.  Therefore
minimization of $W^{\infty}$ corresponds to the experimentally
observed cellular preference for large effective stiffness.

\subsection{Isotropic elastic medium}

The mechanical equilibrium condition Eq.~(\ref{equilibrium1}) states
that the applied body forces $f_i(\vec{r})$ are
balanced by the internal restoring forces $\sigma_{ij,j}(\vec r)$,
where $\sigma_{ij}(\vec r)=C_{ijkl}u_{kl}(\vec r)$ is the stress
tensor.  In the following, we will consider only isotropic elastic
materials, that is there are two elastic constants, e.g.\ the Lam\'{e}
coefficients $\mu$ and $\lambda$ or Young modulus $E$ (\textit{elastic
rigidity}) and Poisson ratio $\nu$ (that describes the relative
importance of shear and compression modes). For our purpose, it is
convenient to define a third pair of elastic constants, $\Lambda =
\lambda / \mu$ and $c = 2 \mu + \lambda = \mu (2+\Lambda)$.  
Therefore Poisson ratio $\nu = \Lambda / 2 (\Lambda + 1)$ and $\nu =
1/2, 1/4$ and $0$ correspond to $\Lambda \rightarrow \infty$, $\Lambda
= 1$ and $\Lambda = 0$, respectively.  In practice, $E$ will be of the
order of a few kPa, which is a typical value for physiological tissues
(simple scaling shows that for a typical force $F = 10$ nN at focal
adhesions, a deformation in the $\mu$m-range corresponds to $E = 10$
kPa). Values for the Poisson ratio $\nu$ are close to $1/2$
(incompressible medium) both for synthetic elastic substrates and
physiological hydrogels.  However, other values for $\nu$ might be
realized in future applications, e.g.\ for artificial tissues or on
compliant surfaces of biosensors. For isotropic elasticity, the
elastic constant tensor of the medium reads 
$C_{ijkl} = \lambda \delta_{ij} \delta_{kl} + 2 \mu \delta_{ik} \delta_{jl}$
and Eq.~(\ref{equilibrium1}) is conveniently rewritten using a vector
notation as:
\begin{equation}  
 \triangle \vec u(\vec{r})+(1+\Lambda) \nabla \nabla \cdot \vec u(\vec{r}) 
= -\frac{\vec f(\vec{r})}{\mu} \qquad \text{$\vec{r}$ in V},
\label{equilibrium3}
\end{equation}
which is a linear second order differential equation for the
displacement field and has to be solved with the appropriate
boundary conditions. 

Since the differential equation Eq.~(\ref{equilibrium3}) is linear,
the superposition principle applies and the boundary value problem is
formally solved by determining the Green tensor
$G_{ij}(\vec{r},\vec{r'})$, i.e.\ the kernel for a point-like body
force $f_i(\vec r)= f_i \delta(\vec{r}-\vec{r'})$.  The elastic fields
of more complicated force distributions can be obtained by convolution
of the Green tensor with the force density, i.e.\ $u_i(\vec{r}) =\int
G_{ij}(\vec{r},\vec{r'}) f_j(\vec{r'}) d^3r'$. The elastic fields
resulting from force dipoles are obtained by differentiation of
$G_{il}$, $u_{i}(\vec{r})= G_{il,k}(\vec{r},\vec{r'}) P_{kl}$ and
$u_{ij}(\vec{r})= G_{il,kj}(\vec{r},\vec{r'}) P_{kl}$. In general, the
determination of Green functions in elasticity theory for a given
geometry and boundary condition is rather difficult, since the second
term in Eq.~(\ref{equilibrium3}) couples different components of the
displacement field. By taking the Laplacian of
Eq.~(\ref{equilibrium3}), one arrives at the biharmonic equation
$\triangle\triangle \vec u=0$ for the displacements.  Thus, harmonic
potential theory is frequently used, for instance in the stress
function $\chi$-method \cite{b:land70} and in the Galerkin-vector
approach \cite{e:mind36}, in addition to other methods like expansion
of $\vec u$ in terms of a suitable complete basis set of orthonormal
functions \cite{e:hirs81}.
 
\subsection{External strain}

We now consider how a cell establishes a force pattern in a
prestrained homogeneous medium. The work required to generate a force
pattern in the presence of an externally imposed strain tensor field
$u_{ij}^e(\vec r)$ is given by:
\begin{eqnarray}
W&=& \int d^3r\ \int_{0}^{u_{ij}^e+u_{ij}^{\vec f}}\
C_{ijkl}u_{kl}(\vec{r})du_{ij}(\vec
r) \nonumber \\
&& - \int d^3r\ \int_0^{u_{ij}^e}\ C_{ijkl}u_{kl}(\vec
r)du_{ij}(\vec{r}) = W^{\infty}+\Delta W^e
\end{eqnarray}
with 
\begin{eqnarray}
\Delta W^e =\int d^3r\ C_{ijkl} u^{\vec f}_{ij}u_{kl}^e(\vec{r}) 
= \sum_{n=0}^{\infty} \frac{1}{n !} P_{i_1 \dots i_n i} u^{e}_{i,i_1
\dots i_n}(\vec{r}).
\label{externalstrain}
\end{eqnarray}
The derivation of Eq.~(\ref{externalstrain}) proceeds along the same
lines as for Eq.~(\ref{referenceEnergy}).  In particular, for a single
force dipole one gets $\Delta W^e = P_{ij}u_{i,j}^e(\vec{r})$.
Because of contractile cell activity, $P_{ij}$ has negative
eigenvalues ($P<0$).  Thus, tensile prestrain ($u_{ij}^e>0$) decreases
$W$ as does a larger rigidity $E$ and hence is interpreted by the cell
as an increase in effective stiffness (\textit{strain-stiffening}). In
contrast, compressive prestrain corresponds to a decrease in effective
stiffness and hence is avoided by the cell.

\subsection{Boundary-induced image strain}

We now consider the energy involved to deform an elastic medium
in the presence of a sample boundary. In order to
quantify the effect introduced by the boundary, we split
$u_{ij}=u_{ij}^{\infty}+u_{ij}^b$ into a contribution arising in an
infinite medium $u_{ij}^{\infty}$ and a boundary induced strain field
$u_{ij}^b$ (\textit{image strain}), that depends on sample geometry
and boundary condition.  $\vec u^{\infty}$ ensures that the force
balance is satisfied everywhere in the sample volume $V$. However,
$\vec u^{\infty}$ will not satisfy the boundary condition at $S$,
that requires to introduce $\vec u^{b}$.  In order to keep the force
balance in the sample, the image displacements have to be homogeneous
solutions of Eq.~(\ref{equilibrium3}).  Otherwise they can be chosen
in such a way that the boundary conditions are satisfied. Now
$W = W^{\infty}+\Delta W^b$, where $W^{\infty}$ is the energy of the
infinite medium and $\Delta W^b$ is the additional energy
due to the boundary effect. For the latter, we have
\begin{equation}
\Delta W^{b}=\frac{1}{2}\int d^3r\ f_i(\vec{r}) 
u^{b}_i(\vec{r})+\frac{1}{2}\oint dS\ f^{s}_i(\vec{r}) u_i(\vec{r})
\label{boundary}
\end{equation}
which includes both the effects of fixed boundary strain and fixed
boundary forces. In principle, the boundary conditions in a
physiological context can be very complicated. In our calculations we
will restrict ourselves to two fundamental reference cases, namely
\textit{free boundaries}, where the normal tractions vanish at the
boundary, i.e.\ $f_i^s(\vec{r})=0$, and \textit{clamped boundaries},
where the displacements vanish at the boundary, i.e.\
$u_i(\vec{r})=0$. We will refer to the former as the \textit{Neumann
problem} and to the later as the \textit{Dirichlet problem}.  In both
cases, the surface integral in Eq.~(\ref{boundary}) vanishes. Thus,
the change in effective stiffness induced by a boundary as encountered
by a force dipole reads $\Delta W^b = \frac{1}{2} P_{ij}
u_{i,j}^b(\vec{r})$.  In this way, cells can actively sense not only
the presence of a close-by surface, but also its shape and boundary
condition.

\subsection{Elastic interactions of cells}

Strain fields produced by other cells may be large enough to be
detected as external strain by the cell, which gives rise to an elastic
interaction of cells.  Even if cells have initially isotropic force
patterns, they will sense anisotropic strain and begin to polarize.
The change in stiffness encountered by a force pattern $\vec f$
centered around $\vec{r}$ caused by a second force pattern $\vec
f^{\prime}$ centered at $\vec{r'}$ reads:
\begin{eqnarray}
\Delta W^{\vec f\vec f^{\prime}} &=& \int d^3 s f_i(\vec{r}+\vec s)u_i(\vec{r}+\vec s)
=\int \int d^3 sd^3s^{\prime} f_i(\vec{r}+\vec s) G_{ij}(\vec{r}+\vec s,\vec{r'}+\vec{s'}) 
f^{\prime}_j(\vec{r'}+\vec{s'}) \nonumber \\
&=&\sum_{n=0}^{\infty}\sum_{m=0}^{\infty}\frac{1}{n!}\frac{1}{m!}
P_{i_1 \dots i_n i} G_{ij,i_1 \dots i_n j_1 \dots j_m}(\vec{r}, \vec{r'}) 
P^{\prime}_{j_1 \dots j_m j}\ ,
\label{elasticInteraction}
\end{eqnarray}
where the indices $i_1$ \dots $i_n$ denote derivatives of the Green function
with respect to the unprimed coordinates and $j_1$ \dots $j_m$ 
derivatives with respect to the primed coordinates.
For translationally invariant geometries, for instance in 
infinite space, $G_{ij}(\vec{r},\vec{r'})=G_{ij}(\vec{r}-\vec{r'})$ and
derivatives for $j_k$ become equivalent to derivatives for $- i_k$.
As a model for elastically interacting cells, we consider how identical, 
static anisotropic contraction dipoles organize in a soft medium 
in order to sense maximal effective stiffness in their environment. 
The case $n = m = 1$ in Eq.~(\ref{elasticInteraction}) corresponds to the 
force dipolar interaction:
\begin{equation} \label{DipoleInteraction}
\Delta W^{PP^{\prime}} = P_{li} u_{i,l}(\vec{r}) 
=  P_{li} G_{ij,lk}(\vec{r},\vec{r'}) P^{\prime}_{kj}
\end{equation}
and will be discussed in more detail below.

\subsection{Summary modeling section}

To summarize the first part of this paper, both physical defects and
active cells respond to elastic deformations in their environment and
we suggest that the same mathematical formalism can be used to
describe both situations.  In fact, all formulae derived in this
section for interactions of cells with external strain, sample
boundaries and other cells as quantified by $W$ describe the
corresponding interactions of physical dipoles as quantified by $V_t$,
with $W$ and $V_t$ being related to each other simply by a switch in
sign.  This result is typical for situations described by energies
with quadratic scaling, as explained above for the simple case of a
linear spring. For different situations of interest we found the same
result $\Delta W = P_{ij} u_{ij}$, where $u_{ij}$ is the strain tensor
evaluated at the position of the force dipole $P_{ij}$.  Depending on
the situation of interest, this strain tensor can correspond to
externally imposed strain, image strain induced by a sample boundary
or strain due to the traction of other force dipoles. Our formula
shows that cells can sense anisotropies in their environment only
through an anisotropic probing process: if the probing process were
isotropic, $P_{ij} = P \delta_{ij}$, we would find $W = P u_{ii}$ and
cells could only sense the scalar quantity $u_{ii}$ describing the
local relative change in volume, but not any tensorial quantity like
the direction of external strain.

It is important to note that the above equations for active cells are
not interaction potentials in a strict physical sense.  Rather these
equations try to quantify information that cells can gain by pulling
on their environment and show how external perturbations result in
changes in effective stiffness. The experimental observation that
active cells prefer large effective stiffness in their environment
leads to the interaction laws for cells given in
Eqs.~(\ref{externalstrain},\ref{boundary},\ref{elasticInteraction}).
In this way, we can predict cellular self-organization in soft media
from an extremum principle in elasticity theory, in excellent
agreement with experimental results \cite{uss:bisc03a}. The structure
formation for physical dipoles follows simply by inverting the sign of
the interaction laws derived for active cells. This case might also
apply to artificial or inert cells \cite{uss:schw02a}. For biomimetic
systems, one might think of vesicles or nanocapsules which contract on
adhesion to an elastic environment. For cellular systems, one might
think of cells which are deficient in regard to the experimentally
observed dynamic response of normal cells to elastic properties of the
environment.

In regard to modeling of active cells, we assume that they probe their
elastic environment through an anisotropic process in which force is
of central importance, and that this process results in a cellular
preference for large effective stiffness in the environment.  Although
the phenomena described here are closely related to cell morphology
and the dynamics of focal adhesions, these aspects are not the subject
of the present work.  In particular, the magnitude $P$ of the cellular
force dipole tensor does not enter our predictions, in contrast to the
positions and orientations represented by the dipole tensor
$P_{ij}$. This reflects the fact that our model focuses on the
extracellular properties that can be sensed by the cell.  Since we
avoid modelling cell morphology and dynamics of focal adhesions, we
are able to describe the active behavior of cells in the same
mathematical framework developed before to describe physical defects
in a deformable medium. In particular, both cases require the solution
of the corresponding elastic boundary value problem given in
Eq.~(\ref{equilibrium1}) and Eq.~(\ref{equilibrium2}). In the next
section, we present exact solutions for different cases of interest.

\section{Examples of cell organization}

\subsection{Interaction with external strain}

As an example for cell organization in a prestrained environment,
we consider a homogeneously prestrained elastic slab with an 
uniaxial stress $p$ acting along the z-axis. The other faces are
traction free, i.e.\ the stress tensor has only one non-zero
component, $\sigma_{zz}=p$ .  Then the corresponding strain tensor has
only diagonal components $u_{ij}^e=\frac{p}{E}\{(-\nu,
0,0),(0,-\nu,0),(0,0,1)\}$ independent of position. Contraction of this
external strain tensor with the force-dipole tensor $P_{ij}$ according
to Eq.~(\ref{externalstrain}) leads to:
\begin{equation}
\Delta W^e=\frac{pP}{E}[(1+\nu) \cos^2 \theta -\nu]\ ,
\label{homogenousStrain}
\end{equation}
where $\theta$ is the orientation of the dipole relative to the
direction of externally applied strain. Eq.~(\ref{homogenousStrain})
applies to both a cell on the top surface of the strained slab
(elastic substrate) or inside a strained infinite elastic material
(hydrogel).  For tensile strain ($p>0$) the cell senses maximal
effective stiffness along the direction of stretch $\theta=0$, thus
cells orient preferentially in the direction of stretch in a
prestrained environment.  On the other hand, due to lateral
contraction, cells in a precompressed environment ($p<0$) will orient
perpendicularly to the axis of compression, which is a combined effect
of compressive strain avoidance in the z-direction and maximal tensile
strain detection in the perpendicular directions, which will be most
pronounced for incompressible media ($\nu\approx 1/2$).  In contrast,
a physical anisotropic contraction dipole, causing a local contraction
of the environment along its axis, is repelled (attracted) by tensile
(compressive) strain, because the negative interaction energy with the
medium is reduced (increased) due to the expansion (compression) of
the environment caused by the external field.  Physical anisotropic
contraction dipoles therefore orient in the opposite way as
mechanosensing cells with respect to external homogeneous strain.

\subsection{Dipoles on elastic halfspace}

Mechanically active cells adhering to an elastic substrate can interact
elastically with each other according to
Eq.~(\ref{elasticInteraction}).  If the thickness of the substrate is
much larger than the elastic displacements on the top surface, it can
be modelled as a semi-infinite elastic space \cite{uss:schw02b}.  The
Green function for a force applied to the surface of a semi-infinite
space is given by the well known Boussinesq solution
\cite{b:land70}. Since tangential forces are expected to be much
larger than normal forces, $P_{ij}$ can be restricted to the
$x$-$y$-plane. Moreover the normal displacement component contributes
very little to the elastic interaction and we may use the
two-dimensional (2D) Green function, i.e.\ only the $x$- and
$y$-components of the Boussinesq solution:
\begin{equation} \label{eq:Green_2d}
G^{2D}_{ij}(\vec{r},\vec{r'}) = a_1 \left\{ 
a_2 \delta_{ij} + \frac{R_i R_j}{R^2} \right\} \frac{1}{R},
\end{equation}
where $\vec{r}=\vec{r}-\vec{r'}$ and
\begin{equation}
a_1 = \frac{\Lambda (\Lambda+2)}{4 \pi c (1+\Lambda)}= \frac{\nu (1+\nu)}{\pi E}\ , \quad
a_2 = \frac{2+\Lambda}{\Lambda}=\frac{1-\nu}{\nu}\ .
\end{equation}
It is convenient to define the angles $\theta$, $\theta^{\prime}$ and
$\alpha$ via the scalar products $\cos \theta = \vec l \cdot \vec{r}$,
$\cos \theta^{\prime}=\vec l^{\prime} \cdot \vec{r}$ and $\cos \alpha
=\vec l \cdot \vec l^{\prime}$.  Then the change in effective
stiffness encountered by one cell due to the traction of the other is
given by:
\begin{equation}
\Delta W^{PP^{\prime}}=a_1 \frac{PP^{\prime}}{2 R^3}f(\theta,\theta^{\prime},\alpha)
\label{2Dinteraction}
\end{equation}
with the angular dependence:
\begin{eqnarray}
f(\theta,\theta^{\prime},\alpha)&=&
3(\cos^2\theta+\cos^2\theta^{\prime}-5\cos^2\theta\cos^2\theta^{\prime}-\frac{1}{3}) \nonumber \\
&-&(1-a_2)\cos^2\alpha -3(a_2-3)\cos\alpha\cos\theta\cos\theta^{\prime}.
\label{2Dangle}
\end{eqnarray}
Since the displacements of a force dipole scale $\sim R^{-2}$, the
strain field scales $\sim R^{-3}$ with distance, which leads to a
long-ranged elastic interaction ($W^{PP^{\prime}} \sim R^{-3}$)
typical for dipolar interactions.  The complicated angular dependence
in Eq.~(\ref{2Dangle}) results in a highly anisotropic
interaction. Note that for the planar geometry, there are only two
independent angles. Nevertheless here we prefer to write the
interaction symmetric in the primed and unprimed coordinates, since
this is favorable for numerical implementations.

$\Delta W^{PP^{\prime}}$ has a pronounced minimum for aligned dipoles
$(\theta=\theta^{\prime}=\alpha=0)$, independent of $\nu$.  A
contractile cell causes a local compression of the substrate
underneath the cell along the contraction axis and tensile strain at
more distant points. Hence at distant points maximal strain-stiffening
occurs along the axis of contraction. A second cell will therefore
upregulate its mechanical activity along the same direction. This
scenario constitutes a positive mechanical feedback loop for cell
alignment, since in the aligned configuration the mechanical activity
of one cell upregulates the activity of the other and vice versa.  At
low cell densities, a common pattern for the organization of
elastically interacting cells will therefore be the formation of
strings of cells, similar to the case of electric dipoles
\cite{e:tlus00}.  Strings might close into rings so that each cell is
fully activated by its neighbors.

\begin{figure}
\begin{center}
\includegraphics[width=0.8\columnwidth]{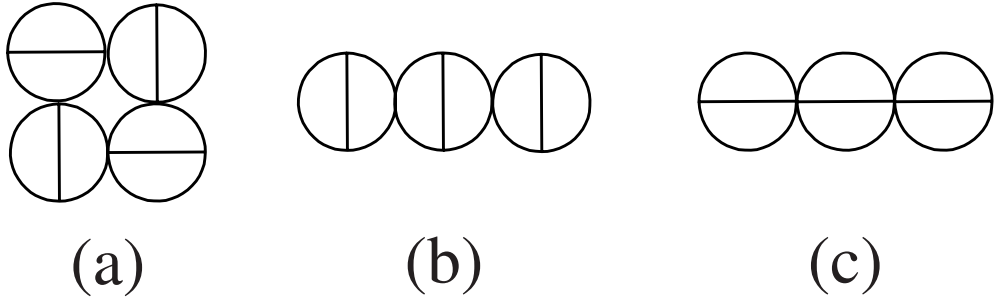}
\caption{\label{fig:structure} Different structures
arising from elastic interactions of anisotropic force dipoles on top
of an elastic halfspace. (a) Physical force dipoles for Poisson ratio
$\nu \approx 1/2$ locally form a T-configuration.  The resulting
structure formation is compact and similar to the one of electric
quadrupoles. (b) Physical force dipoles for Poisson ratio $\nu \approx
0$ align side by side in a railway track like configuration. The
crossover between (a) and (b) occurs at $\nu = 1/5$. (c) Cellular
force dipoles align in strings, similar to electric dipoles and
independent of the value for $\nu$.}
\end{center}
\end{figure}

The 2D case for physical dipoles has been discussed before 
for the isotropic case \cite{e:lau77}. Then 
\begin{equation}
V_t = - P \delta_{li} G_{ij,lk}(\vec{r},\vec{r'}) P' \delta_{kj}  
= - P P' G_{ij,ij}(\vec{r},\vec{r'}) 
= \frac{(2+\Lambda)^2 P P^{\prime}}{4 \pi (1+\Lambda) c R^3}. 
\end{equation}
Thus, for identical dipoles the interaction is isotropic and
repulsive. The case of anisotropic physical dipoles is described by
the negative of Eq.~(\ref{2Dinteraction}).  Then the groundstate
configuration strongly depends on the Poisson ratio $\nu$ via the
angular dependence of Eq.~(\ref{2Dangle}).  For incompressible media,
$\nu=1/2$ ($\Lambda \rightarrow \infty$), dipoles arrange with
perpendicular orientations in a local T-configuration. This leads to
rather compact structure formation, with a square lattice pattern at
intermediate and a herringbone pattern at high dipole densities,
similar to the situation with electric quadrupoles \cite{uss:schw02a}.
For highly compressible media, $\nu\rightarrow 0$ ($\Lambda\rightarrow
0$), dipoles prefer to align side by side in a railway track like
configuration.  For $\nu=1/5$ ($\Lambda = 2/3$), both states have
degenerate energies. \fig{fig:structure} schematically shows the
different structures predicted by our analysis.

\subsection{Dipoles in elastic full space}

Strain propagation in an elastic three-dimensional (3D) infinite
medium is described by the Thomson Green function \cite{b:land70}:
\begin{equation} \label{eq:Green_3d}
G^{\infty}_{ij}(\vec{r},\vec{r'}) = a_1^{\infty} \left\{ 
a_2^{\infty} \delta_{ij} + \frac{R_i R_j}{R^2} \right\} \frac{1}{R},
\end{equation}
with 
\begin{equation}
a_1^{\infty}=\frac{1+\nu}{8 \pi E (1-\nu)} = \frac{\Lambda+1}{8\pi c}\ , \quad
a_2^{\infty}=(3-4 \nu) = \frac{3+\Lambda}{1+\Lambda}\ .
\end{equation} 
The most important result for physical dipoles is the
fact that since $G^{\infty}_{ii} = 0$, the elastic interaction of
isotropic dipoles in 3D vanishes \cite{e:siem68}.  Therefore their
interaction is completely determined by boundary-induced interactions,
like for hydrogen in metal samples of finite size \cite{e:wagn74}.

\begin{figure}
\begin{center}
\includegraphics[width=0.8\columnwidth]{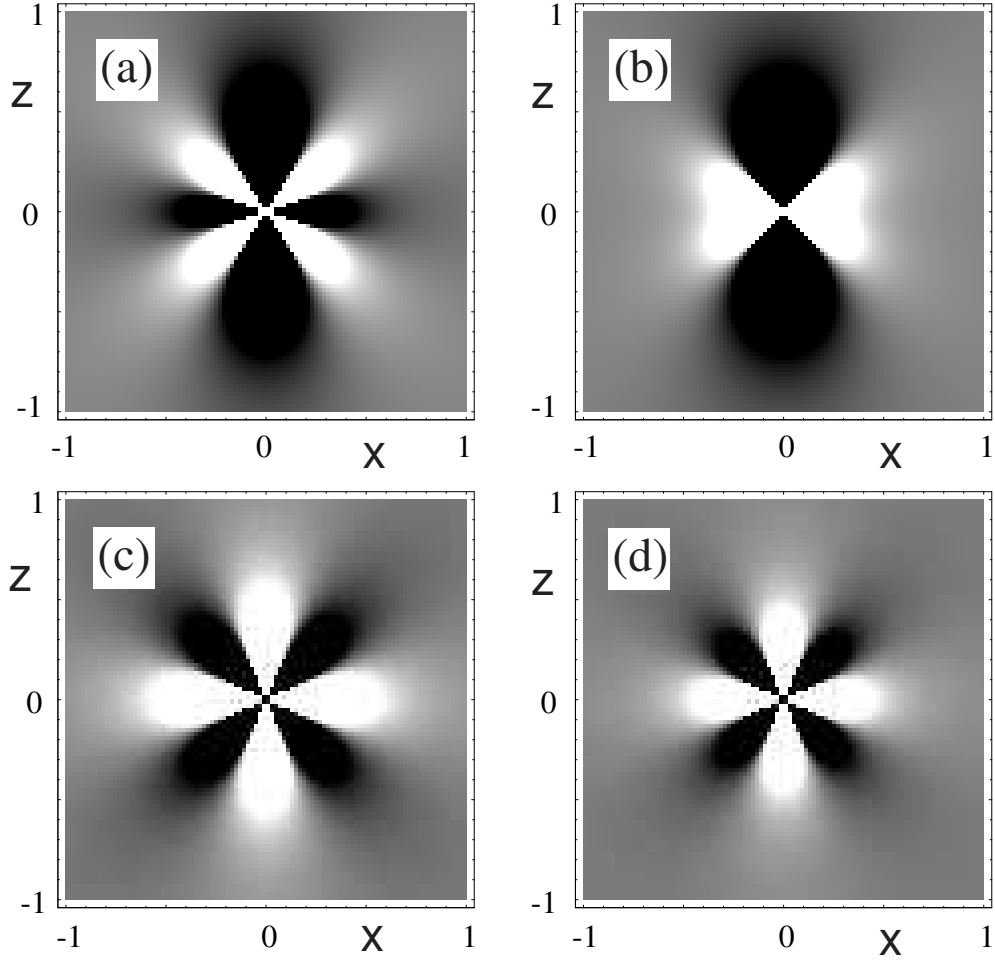}
\caption{\label{fig:3Dinteract} Density plots of cellular interaction 
potential $\Delta W^{PP'}$ from \eq{3Dinteract} for (a,b) parallel and
(c,d) perpendicular orientations. In (a,c), Poisson ratio $\nu = 1/2$,
and in (b,d), $\nu = 0$. One dipole oriented along the z-axis is fixed
at the origin, while the other is moved in space. Black denotes areas
of attraction and white areas of repulsion. The interaction potential
for physical force dipoles simply differs in sign, thus black and
white exchange meaning. (a,b) Independent of the value for $\nu$, two
cells prefer alignment (black region along z-axis).  The interaction
in the railway track configuration (along x-axis) changes sign at $\nu
= 1/4$, when the black cone vanishes.  (c,d) The T-configuration is
the ground state for physical dipoles in 3D independent of the value
for $\nu$ (white regions along z- and x-axes). This is different on an
elastic halfspace, in which case the groundstate changes from the T-
to the railway track configuration for $\nu = 1/5$.}
\end{center}
\end{figure}

For the elastic interaction of two active cells, we find
\begin{equation}
\Delta W^{PP^{\prime}}=a_1^{\infty}\frac{PP^{\prime}}{2 R^3}
f^{\infty}(\theta,\theta^{\prime},\alpha)
\label{3Dinteract}
\end{equation}
with the angular function $f^{\infty}(\theta,\theta^{\prime},\alpha)$
given by Eq.~(\ref{2Dangle}) by replacing the constants $a_1$ and
$a_2$ with $a_1^{\infty}$ and $a_2^{\infty}$, respectively. Note that
in 3D there are three independent orientational degrees of freedom.
In \fig{fig:3Dinteract} we show a density plot of $\Delta W^{PP^{\prime}}$
for dipoles with relative orientations $\alpha=0$ and $\alpha = \pi/2$
positioned in the x-z-plane for two different values of the Poisson
ratio, $\nu=0$ and $\nu=1/2$.  Like on 2D substrates, cells in a 3D
environment encounter a mechanical feedback loop favoring cell
alignment.  For two parallel dipoles in z-direction placed along the
z-axis, we find
\begin{equation}
\Delta W^{PP^{\prime}} = -\frac{(\Lambda + 2) P^2}{2 \pi c} 
\left( \frac{1}{z} \right)^3,
\label{3DW}
\end{equation}
which yields the optimal configuration independent of the value for
$\Lambda$ (or, equivalently, $\nu$). Again this behaviour is similar
to the ones of electric dipoles \cite{e:tlus00}.  For two parallel
dipoles in z-direction placed along the x-axis (railway track
configuration), we find
\begin{equation} \label{direct}
\Delta W^{PP^{\prime}} = \frac{(\Lambda - 1) P^2}{8 \pi c} \left( \frac{1}{x} \right)^3\ .
\end{equation}
Thus $\Delta W^{PP^{\prime}}$ changes sign as $\Lambda$ varies through $1$ ($\nu =
1/4$).  Finally, in the T-configuration, where the first dipole is
fixed in z-direction at the origin and the second dipole is positioned
in the x-y-plane oriented perpendicular to the z-axis, we find:
\begin{equation}
\Delta W^{PP^{\prime}} = - \frac{(\Lambda + 1) P^2}{4 \pi c} 
\left( \frac{1}{r} \right)^3
\end{equation}
which is always positive and yields a globally maximal $\Delta
W^{PP^{\prime}}$.  Therefore it corresponds to a globally minimal $V_t
= - \Delta W^{PP^{\prime}}$ and the T-configuration is the groundstate
of two physical anisotropic contraction dipoles, independent of the
value for $\nu$.  The aggregation of physical dipoles in 3D is more
complicated than in 2D, since the T-configuration cannot be continued
in 3D without causing frustration. This leads to the existence of many
metastable states.

\subsection{Dipoles in elastic halfspace}

The elastic isotropic halfspace with a clamped surface at $r_3=0$
constitutes a Dirichlet problem with vanishing displacements at the
planar boundary, $u_i(\vec{r})=0$ for $r_3=0$, whereas the free
surface leads to a Neumann boundary value problem with vanishing
surface tractions, $\sigma_{ij} (\vec{r}) n_j=0$ for $r_3=0$ with
$\vec n=(0,0,1)$ being the surface normal.  The boundary value problem
of the semi-infinite space can be solved using the concept of image
singularities. Image approaches are well known from electrostatics:
the simplest example is the charge in front of a metal plate. Here,
the field due to a charge $Q$ at $\vec
r^{\prime}=(r_1^{\prime},r_2^{\prime},r_3^{\prime})$ with the boundary
at $r_3=0$ is equivalent to the field of the charge and an image
charge $-Q$ at $\vec{r'}_{\rm
im}=(r_1^{\prime},r_2^{\prime},-r_3^{\prime})$ without a boundary. In
analogy, the displacement field due to a unit force at $\vec{r'}$
close to a planar surface of a semi-infinite space is equivalent to
the superimposed fields of a set of force nuclei placed in a
homogeneous infinite substrate, i.e.:
\begin{equation}
G_{ij}(\vec{r},\vec{r'})=G_{ij}^{\infty}(\vec{r},\vec{r'})+
G_{ij}^{\rm im}(\vec{r},\vec{r'}),
\end{equation}
where $G_{ij}^{\infty}$ is the Green function in an infinite medium,
Eq.~(\ref{eq:Green_3d}), and $G_{ij}^{\rm im}$ specifies its image
system, which is a sum of functions derived from $G_{ij}^{\infty}$ by
differentiation (point images, i.e.\ forces and force dipoles)  or
integration (line images, i.e.\ a line of force nuclei).  Despite its
rather simple geometry, the image system of the elastic half-space is
rather complicated and consists of up to 15 image nuclei,
including point nuclei located at $\mathbf r^{\prime}_{\rm
im}=(r_1^{\prime},r_2^{\prime},-r_3^{\prime})$ and line images running
normal to the surface and extending from $-r_3^{\prime}$ to infinity.
The image system of the free halfspace was calculated by Mindlin using
a Boussinesq-Galerkin representation \cite{e:mind36}. The Green
function of the clamped half-space has been derived by Phan-Thien
applying a Papkovitch-Neuber ansatz, however without revealing the
image system in detail \cite{e:phan83}.  Quite recently, Walpole
\cite{e:walp96} used methods of general harmonic potential theory and
presented the image system for two joined half-spaces, which includes
the clamped or free half-space as limiting cases of infinite or
vanishing shear rigidity in one of the joined spaces.  Introducing the
harmonic functions:
\begin{eqnarray}
\frac{1}{s}&=&\frac{1}{|\vec{r}-\vec{r'}_{\rm im}|},
\end{eqnarray}
where $s$ the distance from the image point, and
\begin{eqnarray}
\Phi&=&\ln (r_3+r_{3'}+s) \nonumber \\
\Psi&=& (r_3+r_3^{\prime}) \Phi-s,
\end{eqnarray}
the image Green tensor $G_{ij}^{\rm im}$ of the isotropic elastic 
half-space reads \cite{e:walp96}:
\begin{eqnarray}
G_{ij}^{\rm im}(\vec{r}, \vec{r'})&=& M G_{ij}^{\infty}(\vec{r},\vec{r'}_{\rm im})+ \nonumber \\
&+&\frac{J r_3^{\prime}(1+\nu)}{4 \pi E (1-\nu)} \left[s_{,ij3}-2\delta_{j3} s_{,i33}-4 (1-\nu) \delta_{i3}\left[\left(\frac{1}{s}\right)_{,j}-2 \delta_{j3}\left(\frac{1}{s}\right)_{,3}\right]\right] -\nonumber \\
&-& \frac{J r_3^{\prime}(1-2\nu)(1+\nu)}{2 \pi E (1-\nu)} \delta_{j3} \left(\frac{1}{s}\right)_{,i}- \nonumber \\
&-& \frac{J r_3^{\prime 2}(1+\nu)}{4\pi E (1-\nu)}  \left[\left(\frac{1}{s}\right)_{,ij}-2\delta_{j3}\left(\frac{1}{s}\right)_{,i3}\right]- \nonumber \\
&-&\frac{C (1-2\nu)(1+\nu)}{4\pi E (1-\nu)} (\Psi_{,ij}-2 \delta_{j3}
\Psi_{,i3})+ \frac{B(1+\nu)}{2 \pi E} \delta_{j3} \Phi_{,i}+ \nonumber \\
&+&\frac{B (1+\nu)}{2\pi E} (\delta_{i3} \Phi_{,j}-\delta_{ij} \Phi_{,3})),
\label{HalfspaceGF}
\end{eqnarray}
where the coefficients $M,J,C,B$ depend on the boundary condition
(subscripts: free $f$, clamped $c$) and the Poisson ratio $\nu$ \cite{e:walp96}:
\begin{eqnarray}
M^{f}= (3-4\nu)\qquad  & M^{c}=-1 \nonumber \\
J^{f}= -1\qquad  & J^{c}= 1 / (3-4\nu) \nonumber \\
C^{f}= 2 (1-\nu)\qquad & C^{c}=0 \nonumber \\
B^{f}= 2 (1-2\nu)\qquad & B^{c}=0.
\end{eqnarray}
For a fixed $j$, each line in Eq.~(\ref{HalfspaceGF}) represents the
$i$-th component of the displacement field of one fundamental strain
nuclei of an infinite medium. For a free surface, five image
singularities contribute to a surface tangential or normal force
component. A tangential force $j=1,2$ introduces, in the order of
lines of Eq.~(\ref{HalfspaceGF}), three point images (force, double
force with moment and a doublet) and two line images (line of doublets
and line of double forces with moment) \cite{e:mind36}.  A normal
force $j=3$ induces four point images (force, double force, doublet,
center of compression/dilation) and a line of compression/dilation
centers \cite{e:mind36}.  In a clamped halfspace the line images
disappear ($B=C=0$) and there are only the three or four point images
for a tangential or normal force component, respectively.
Interestingly, the strength of the higher order point singularities is
proportional to the distance $r_3^{\prime}$ of the source point from
the surface. Hence their relative contribution to the displacement
field with respect to the image force increases with increasing
distance of the source force from the surface.  Note that for
$r_3^{\prime}\rightarrow 0$, i.e.\ for a force acting at a free surface
of a semi-infinite space, Eq.~(\ref{HalfspaceGF}) yields the Boussinesq
Green function from Eq.~(\ref{eq:Green_2d}) for tangentially applied forces
and the solution of Cerruti for normally applied forces.  The dominant
terms to the image displacement field far away from the surface arise
from the image force and the line images $\sim 1/s$, followed by the
dipole type defects (double force, compression center) $\sim
r_3^{\prime}/s^2$ and finally the doublet $\sim r_3^{\prime 2}/s^3$.
The Poisson ratio $\nu$ changes the relative magnitude of the image
singularities with respect to each other, but does not change their
type (i.e. their sign).  Therefore, strain propagation in the
halfspace is expected to stay qualitatively similar with
varying $\nu$. Changing the boundary condition from free to clamped, the point
images flip their sign, which indicates that clamped and free boundary
will induce qualitatively opposite effects. Indeed, for the special
case of an incompressible medium, $\nu=1/2$, clamped and free halfspace
induce the same boundary fields but with opposing signs.

\begin{figure}
\begin{center}
\includegraphics[width=\columnwidth]{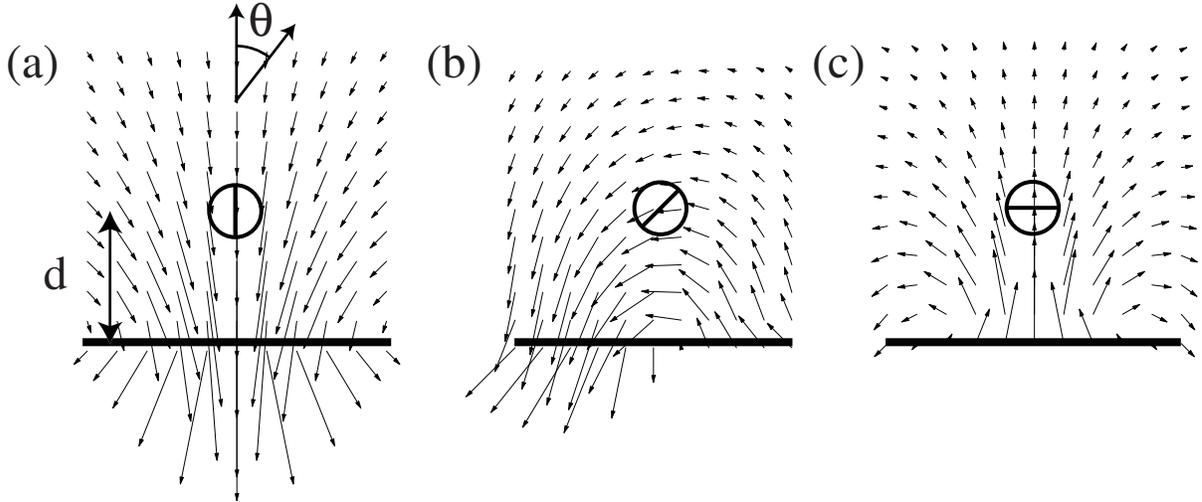}
\caption{\label{fig:Halfspace} Image fields $\vec u^b$ for a contraction dipole 
$P_{ij}$ positioned at $\vec{r'}=(0,0,d)$ in front of a clamped
surface of a semi-infinite space for Poisson ratio $\nu=1/2$.  Dipole
orientations are (a) $\theta=0$, (b) $\theta=\frac{\pi}{4}$ and (c)
$\theta=\frac{\pi}{2}$ with respect to the surface normal.  At the
clamped surface the image displacements $\vec u^b$ balance the
displacements $\vec u^{\infty}$ of an infinite space.  Inside the
sample, they are homogeneous solutions of the elastic equations. The
interaction of a dipole with the clamped surface is equivalent to the
interaction of the dipole with a set of image singularities placed at
$\vec{r'}_{\rm im}=(0,0,-d)$.  For a free surface, the normal
tractions vanish and all image displacements change sign. For $\nu <
1/2$, there is an additional contribution to $\vec u^b$ derived from
line images.  However, the interaction of force dipoles with the
boundary does not change qualitatively as $\nu$ is varied.}
\end{center}
\end{figure}

The image displacements $\vec u^b$ induced by a force dipole $P_{ij}$ at
$\vec{r'}$ are obtained from Eq.~(\ref{HalfspaceGF}) by
differentiation with respect to the primed coordinates. Note that the
planar surface at $r_3=0$ breaks the translational invariance along
the z-axis, which means that differentiation of $G_{ij}^b$ with
respect to $r_3$ and $r_3^{\prime}$ are not equivalent.  Since the
strength of the dipolar singularities in $G_{ij}^{\rm im}$ is
proportional to $r_3^{\prime}$, taking the derivative with respect to
$r_3^{\prime}$ will lead to dipole images of
$r_3^{\prime}$-independent strength that are proportional to the
dipole strength $P$.  Therefore, the far field image displacements
produced by a force dipole in front of a planar surface are dominated
by image dipole terms $\sim 1/s^2$ of strength proportional to $M$ and
$J$ and additional images derived from the line image terms.  In
\fig{fig:Halfspace} we plot $\vec u^b$ for three different dipole
orientations with respect to the surface normal of a clamped halfspace
for Poisson ratio $\nu=1/2$. In this case, all image displacements
point in the opposite direction for a free surface.

\begin{figure}
\begin{center}
\includegraphics[width=0.4\columnwidth]{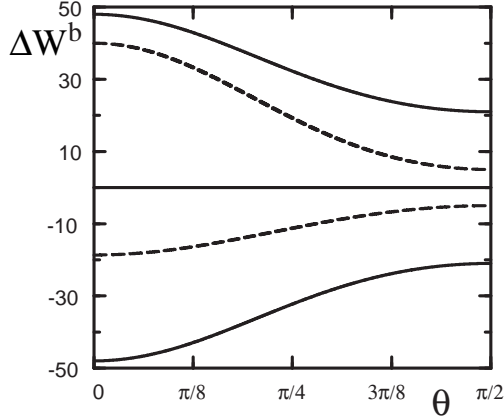}
\caption{\label{fig:HalfspacePNAS} Angular dependence of image interaction
with the boundary, $\Delta W^b$ from \eq{Halfspace}, for a cellular
force dipole positioned a distance $d$ away from the surface of an
elastic halfspace, plotted in units of $P^2/Ed^3$ and rescaled by
$1/256 \pi$. Curves above and below the $\theta$-axis correspond to
free and clamped boundaries, respectively. Solid and dashed lines
correspond to $\nu=1/2$ and $\nu=0$, respectively (all other Poisson
ratios yield curves lying in between those shown). A clamped (free)
surface effectively rigidifies (softens) the medium towards the
surface. Hence, irrespective of the value of $\nu$, cells close to a
clamped surface prefer to orient perpendicular ($\Delta W^b$ minmal
for $\theta = 0$) while cells close to a free surface prefer parallel
orientation ($\Delta W^b$ minmal for $\theta = \pi/2$).}
\end{center}
\end{figure}

According to Eq.~(\ref{boundary}), the change in effective stiffness
encountered by a force dipole $P_{ij}$ positioned a distance $r^{\prime}_3=d$
away from the surface is proportional to the induced image strain at
the position of the dipole, i.e. $\Delta W^b(\vec{r'})=
\frac{1}{2} P_{ij}  \frac{\partial^2 G_{ik}^{\rm im}(\vec{r}, \vec{r'})}
{\partial r_j \partial r^{\prime}_l} P_{kl} |_{\vec{r}\rightarrow \vec
r^{\prime}}$.  Because of rotational symmetry with respect to the
surface normal, the surface induced change in effective stiffness
sensed by a dipole depends only on its distance $d$ to the surface and
the angle $\cos\theta =\vec n \cdot \vec l$ between dipole orientation
and surface normal. We find:
\begin{eqnarray}
\Delta W^b(d, \theta)=\frac{P^2}{256 \pi E d^3} (a_{\nu}+b_{\nu} \cos^2\theta+c_{\nu} \cos^4\theta),
\label{Halfspace}
\end{eqnarray}
with the coefficients
\begin{eqnarray}
a_{\nu}^f=\frac{(1+\nu)(5+2 \nu(6\nu-1))}{1-\nu} \qquad &a_{\nu}^c&=-\frac{(1+\nu)(15+32 \nu (\nu-1))}{(1-\nu)(3-4\nu)} \nonumber  \\
b_{\nu}^f=\frac{(1+\nu)(22+4 \nu(2 \nu-9))}{1-\nu}\qquad &b_{\nu}^c&=-\frac{(1+\nu)(34+32 \nu^2-72 \nu)}{(1-v)(3-4\nu)}  \nonumber \\
c_{\nu}^f=\frac{(1+\nu)(13(1-2 \nu)+12\nu^2)}{1-\nu} \qquad &c_{\nu}^c&=-\frac{(1+\nu)(7-8 \nu)}{(1-\nu) (3-4\nu)}
\label{HalfspaceCoeff}
\end{eqnarray}
being rational function of the Poisson ratio $\nu$.  $\Delta W^b$
scales quadratically in $P$, because the image strain scales linearly
in $P$, in other words, the force dipole interacts with its own
images.  The interaction of the force dipole with the surface is a
long-ranged effect and scales like a dipole-dipole interaction
potential, that is $\sim d^{-3}$ .  For free and clamped surfaces, all
coefficients in Eq.~(\ref{HalfspaceCoeff}) are positive and negative,
respectively, irrespective of $\nu$.  Therefore, the prefered cell
orientation close to the surface , i.e.\ the configurations of minimal
$\Delta W^b$, are parallel ($\theta = \pi/2$) and perpendicular
($\theta= 0$) orientation for free and clamped boundaries,
respectively. In \fig{fig:HalfspacePNAS} we plot the angular
dependence of $\Delta W^b$ for $\nu=1/2$ and $\nu=0$.

Since $|\Delta W^b|\sim 1/d^3$ increases if $d$ decreases, the overall
mechanical activity of a cell increases towards a clamped surface
($\Delta W < 0$), but decreases towards a free surface ($\Delta W >
0$). Thus we predict that cells preferentially locomote towards a
clamped boundary, but tend to migrate away from a free boundary. In
general, free and clamped boundaries have always opposite effects. One
may think of a clamped (free) surface as the interface between the
medium and an imaginary medium of infinite (vanishing) rigidity, which
effectively rigidifies (softens) the medium towards the boundary.
Thus for clamped (free) boundary conditions, the cell senses maximal
stiffness towards (away) from the boundary.  For clamped boundaries,
mechanical activity of cells is favored and cells can amplify this
effect by adjusting orientation. For free boundaries, mechanical
activity of cells is disfavored and the orientation response is an
aversion response.

For the interaction of a physical dipole with the surface, we simply
have to switch sign in Eq.~(\ref{Halfspace}).  Hence, physical dipoles
are attracted by free and repelled from clamped surfaces. A clamped
surface prevents the defect from displacing its environment to lower
its potential energy, which results in a repulsive interaction. In
contrast a free surface favors displacements close to the surface
since at a free surface there exist no internal restoring forces
acting normal to the surface. This results in an attractive
interaction of the defect with the surface.  Since $V_t \sim P^2$, the
sign of $P$ does not matter, i.e.\ dilation and contraction dipole
interact in the same way with the surface.

\subsection{Dipoles in elastic sphere}

As an example for a finite sized sample, we consider the elastic
sphere with radius $R$.  For the elastic sphere, no image system has
been constructed that solves the elastic boundary value problem and it
is not clear whether such an image system exists.  Nevertheless, the
elastic equations for the elastic sphere can be solved analytically by
applying an expansion in terms of vector spherical harmonics.  This
approach has been used by Hirsekorn and Siems \cite{e:hirs81} to solve
the Neumann problem of an anisotropic force dipole in an elastic
sphere with a free boundary. We will follow this approach also in
order to solve the Dirichlet problem of a force dipole in a clamped
sphere.  Both results are then used to calculate the change in
effective stiffness encountered by a force dipole in clamped and
free spheres, respectively.

Analytical solutions to differential equations for scalar fields in
spherical coordinates can be obtained by an expansion in terms of
spherical harmonics, which form a complete orthonormal basis set on
the unit sphere.  In a similar way, the general solution to the
equilibrium condition Eq.~(\ref{equilibrium3}) for the vector field
$\vec u(\vec{r})$ can be expressed as a sum over so-called
\textit{vector spherical harmonics} (VSH):
\begin{equation}
\vec {u}(r,\Omega)=\sum_{lm} f_{lm}(r) \mathbf{Y}^{\dagger}_{ll+1m}(\Omega)
+g_{lm}(r) \mathbf{Y}^{\dagger}_{ll-1m}(\Omega)+h_{lm}(r)\mathbf{Y}^{\dagger}_{llm}(\Omega).
\label{ansatz}
\end{equation}
Vector spherical harmonics $\mathbf{Y}_{JLM}(\Omega)$ form a complete 
orthonormal basis set on the unit sphere:
\begin{equation}
\int \mathbf{Y_{JLM}}(\Omega)\mathbf{Y^{\dagger}_{J^{\prime}L^{\prime}M^{\prime}}}(\Omega)d\Omega 
=\delta_{JJ^{\prime}}\delta_{LL^{\prime}}\delta_{MM^{\prime}}.
\label{orthogonality}
\end{equation}
They are the eigenfunctions of the angular momentum operator $\bf J$
of a vector field as spherical harmonics $Y_{lm}$ are the
eigenfunctions of the (orbital) angular momentum $L$ of a scalar
field.  $\bf J$ is the vector sum $\bf J=L+S$ of the orbital momentum
$\bf L$ and the intrinsic spin $\bf S$. The eigenvectors of $\bf S$
are the spherical basis vectors $\bf e_{\alpha}$:
\begin{eqnarray}
\mathbf{e}_{\pm1}=-\frac{1}{\sqrt2} (\mathbf {e}_x \pm \mathbf {e}_y)\ , 
\qquad
\mathbf{e}_0=\mathbf{e}_z
\label{sphericalbasis}
\end{eqnarray} 
and represent a spin $S=1$ system.  Since $\bf J$ is an example of
angular momentum addition, one can construct the VSH with the help of
Clebsch Gordon coefficients $C^l_{M-\alpha}\,^1_{\alpha}\,^J_ M$
\cite{b:edmo74}:
\begin{equation}
\mathbf{Y}_{JlM}(\Omega)=\sum_{\alpha} C^l_{M-\alpha}\,^1_{\alpha}\,^J_ M  Y_{lM-\alpha}(\Omega) \mathbf {e}_{\alpha}. 
\end{equation}
This implies that for a given $J$ there are only three classes of VSH,  
namely $l=J,J\pm 1$, which in retrospective justifies our ansatz 
Eq.~(\ref{ansatz}).

In order to solve the boundary value problem, we split $\vec u$ again into a 
contribution in an infinite substrate $\vec u^{\infty}$ and a boundary induced 
field $\vec u^b$. $\vec u^{\infty}$ is the solution to the inhomogenous 
differential equation Eq.~(\ref{equilibrium3}) with a body force density
and thus ensures force balance everywhere inside the sample.
For a force dipole $P^{\prime}$ located at $\vec{r'}$ 
the VSH-expansion of the displacement field $\vec u^{\infty}(\vec{r})$ 
reads for $r^{\prime}<r$  \cite{e:hirs81}:
\begin{eqnarray}
\vec u^{\infty}(\vec{r})&=&
\frac{1}{c}\sum_{lm}\frac{\mathbf{Y}^{\dagger}_{ll+1m}(\Omega)}{(2l+1)r^2}
X_{lm}^{\alpha\beta}(\eta^{\prime},\Omega^{\prime})P^{\prime}\,_{\alpha}\,^{\beta}- 
\nonumber \\
&-&\frac{1}{c}\sum_{lm}\frac{\mathbf{Y}^{\dagger}_{ll-1m}(\Omega)}{(2l+1)r^2}
(3l+2+(l+1)\Lambda)C_{m-\alpha}^{l-1}\,_\alpha^1 \,_m^l
A_{l-2 m}^{\alpha \beta}(\Omega^{\prime})\eta^{\prime l-2}P^{\prime}\,_{\alpha}\,^{\beta}- \nonumber \\
&-&\frac{1}{c}\sum_{lm}\frac{\mathbf{Y}^{\dagger}_{llm}(\Omega)}{r^2}(2+\Lambda)
C_{m-\alpha}^l \, _\alpha ^1 \,_m^l A^{\alpha \beta}_{l-1m}(\Omega^{\prime})\eta^{\prime l-1}P^{\prime}\,_{\alpha}\,^{\beta},
\label{VSHuinfty}
\end{eqnarray}
where $\eta^{\prime}=\frac{r^{\prime}}{r}<1$ and
\begin{eqnarray}
A_{lm}^{\alpha \beta}(\Omega)&=&\sqrt{\frac{l+1}{2l+1}} C^{l+1}_{m-\alpha}\, ^{1}_{\beta}\,^l_{l-\alpha+\beta} Y_{lm-\alpha+\beta}(\Omega) \nonumber \\
B_{lm}^{\alpha \beta}(\Omega)&=&\sqrt{\frac{l}{2l+1}} C^{l-1}_{m-\alpha}\,^1_{\beta}\,^l_{m-\alpha+\beta}Y_{lm-\alpha+\beta}(\Omega) \\
X_{lm}^{\alpha \beta}(r,\Omega)&=&-(3l+1+l\Lambda) C^{l+1}_{m-\alpha}\,^1_{\alpha}\,^l_m A^{\alpha \beta}_{lm}(\Omega) r^l 
+\sqrt{l(l+1)}(1+\Lambda) C^{l-1}_{m-\alpha}\,^1_{\alpha}\,^l_m \nonumber \\
&\cdot& [B^{\alpha \beta}_{lm}(\Omega)r^l+\frac{1}{2}
A^{\alpha \beta}_{l-2m}(\Omega)r^{l-2}((2l-1)-(2l+1)r^2)] \nonumber.
\end{eqnarray}
Sums over repeated indices are always implied except for
Clebsch-Gordon coefficients.  $P^\prime\,_{\alpha}\,^{\beta}$ is the
force dipole tensor in the spherical basis set given by
Eq.~(\ref{sphericalbasis}). The reciprocal basis vectors are $\bf
e^{\alpha}=e_{\alpha}^{\dagger}=(-1)^{\alpha} e_{-\alpha}$ and the
metric tensor is $g_{\alpha \beta}=(-1)^{\beta}
\delta_{\alpha,-\beta}$.  Spherical coordinates transform via the
unitary operator $U_{\alpha i}=(\mathbf{e_{\alpha}}\cdot\mathbf{
e}_i)$ into cartesian coordinates, i.e.
\begin{equation}
P_{ij}=U_{\alpha i}U^{\beta}\,_jP^{\alpha}\,_{\beta}.
\end{equation}

In order to satisfy force balance inside the sphere volume,
the boundary induced field $\vec u^b$ must be a homogenous solution 
to Eq.~(\ref{equilibrium3}).
Thus, inserting Eq.~(\ref{ansatz}) into Eq.~(\ref{equilibrium3}), 
one obtains a set of differential equations for the radial functions 
$f_{lm}(r)$, $g_{lm}(r)$ and $h_{lm}(r)$ of the boundary induced field
 \cite{e:hirs81}:
\begin{eqnarray}
0&=&(3l+2+(l+1)\Lambda)(f_{lm}^{\prime\prime}+\frac{2}{r}f_{lm'}-\frac{(l+1)(l+2)}{r^2}f_{lm})-\nonumber\\
&-&\sqrt{l(l+1)}(1+\Lambda)(g_{lm}^{\prime\prime}-\frac {2l-1}{r}g_{lm'}+\frac{(l-1)(l+1)}{r^2}g_{lm})
\label{DGLset1}
\\
0&=&(3l+1+l\Lambda)(g_{lm}^{\prime\prime}+\frac{2}{r}g_{lm'}-\frac{l(l-1)}{r^2}g_{lm})-
\nonumber \\
&-&\sqrt{l(l+1)}(1+\Lambda)(f_{lm}^{\prime\prime}+\frac{(2l+3)}{r}f_{lm'}+\frac{l(l+2)}{r^2}f_{lm})
\\
0&=&h_{lm}^{\prime\prime}+\frac{2}{r}h_{lm'}-\frac {l(l+1)}{r^2}h_{lm}.
\label{DGLset3}
\end{eqnarray}
The general solution to Eq.~(\ref{DGLset1})-Eq.~(\ref{DGLset3}) 
with a $\vec u^b$ which is analytic at the sphere origin is
\cite{e:hirs81}:
\begin{eqnarray}
f_{lm}(r)&=&a_{lm} \frac{3l+1+l\Lambda}{(1+\Lambda)(2 l+3)} r^{l+1}\\
\nonumber \\ \nonumber g_{lm}(r)&=&a_{lm} \frac{1}{2} \sqrt{l (l+1)}
r^{l-1}(r^2-R^2)+b_{lm} \frac{1}{2}r^{l-1} \\ \nonumber \\ \nonumber
h_{lm}(r)&=&c_{lm} r^l,
\end{eqnarray}
where $R$ is the radius of the sphere and the remaining constants
$a_{lm}$, $b_{lm}$ and $c_{lm}$ must be determined by the boundary
conditions at the sphere surface.

\begin{figure}
\begin{center}
\includegraphics[width=0.8\columnwidth]{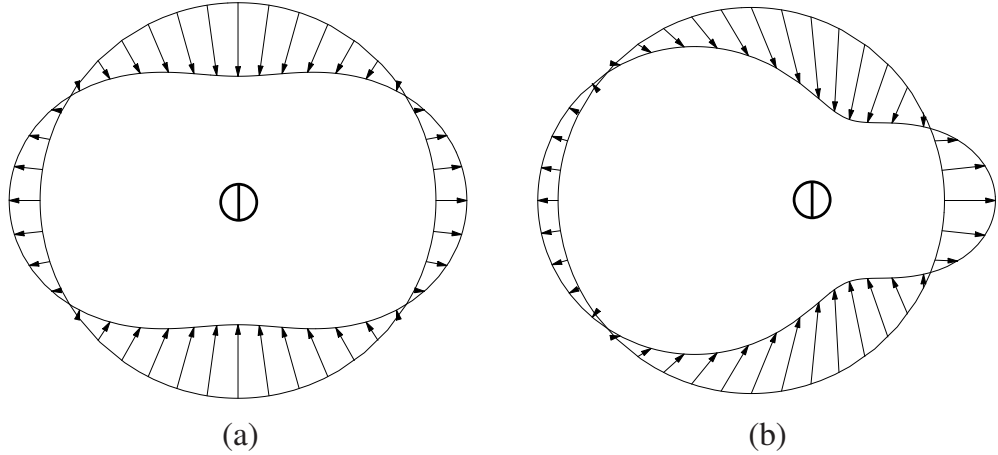}
\caption{\label{fig:SphereInteract} Deformation of an elastic sphere 
($R=1$, $\Lambda=2$, $c=1$) with a free surface by a contraction
dipole oriented in the z-direction.  In (a) the dipole is placed at
the origin, $\vec{r}=(0,0,0)$.  In (b) the dipole is placed off-center
at $\vec{r}=(\frac{R}{4}, 0, 0)$.  The pictures show a cut through the
x-z-plane, but it has rotational symmetry only in (a).}
\end{center}
\end{figure}

The Dirichlet problem of a clamped sphere yields:
\begin{eqnarray}
\vec u^b(R,\Omega)= - \vec u^{\infty}(R,\Omega),  
\label{clampedboundary}
\end{eqnarray}
i.e.\ the expansion coefficients $a_{lm}^c$ etc. of the boundary induced field 
can be found by matching $\vec u^{\infty}$ and $\vec u^b$ at the sphere 
surface:
\begin{eqnarray}
a_{lm}^c&=&-\frac{1}{c R^3} \frac{(2 l+3) (1+\Lambda)}{(2 l+1)(3l+1+l\Lambda) R^l} X_{lm}^{\gamma \delta}(\rho^{\prime},\Omega^{\prime}) P^{\prime}\medspace_{\gamma}\medspace ^{ \delta} \nonumber \\
b_{lm}^c&=& \frac{2}{c R^3} \frac{3 l+2+(l+1)\Lambda}{2l+1} \left( \frac{\rho^{\prime}}{R}\right)^{l-2}  C^{l-1}_{m-\gamma}\,^1_{\gamma}\,^l_m  A^{\gamma \delta}_{l-2 m}(\Omega^{\prime}) P^{\prime}\medspace_{\gamma}\,^{\delta} \\ 
c_{lm}^c&=&\frac{1}{cR^3} (2+\Lambda) \left (\frac{\rho^{\prime}}{R}\right)^{l-1}  C^l_{m-\gamma}\,^1_{\gamma}\,^l_m A^{\gamma \delta}_{l-1 m}(\Omega^{\prime}) P^{\prime}\thinspace_{\gamma}\thinspace^{ \delta}  \nonumber,
\label{clamped-sphere}
\end{eqnarray}
where $\rho^{\prime}=r^{\prime}/R$ is the ratio of the distance
$r^{\prime}$ of $P^{\prime}$ to the sphere center and the sphere
radius $R$.  For a sphere with a free surface normal stress has to
vanish and the corresponding Neumann boundary condition reads:
\begin{equation}
\sigma_{ij}^b(\frac{x_j}{r})_{r=R}=-\sigma_{ij}^{\infty}(\frac{x_j}{r})_{r=R}.
\end{equation}          
To determine $a^f_{lm}$ etc. one first has to calculate the
stress-tensor $\sigma^{\infty}_{ij}$ and then balance the normal
stress with the corresponding boundary induced stress
$\sigma_{ij}^{b}$ at $r=R$. The final result for the expansion
coefficients in a free sphere is \cite{e:hirs81}:
\begin{eqnarray}
a^f_{lm}&=&\frac{1}{cR^3} \frac{2 (1+\Lambda) (2l+3) (l+2)}{(2l+1) M(l) R^l} X_{lm}^{\gamma \delta}(\rho^{\prime},\Omega^{\prime}) P^{\prime}\medspace_{\gamma}\medspace ^{ \delta} \nonumber \\
b^f_{lm}&=&-\frac{1}{cR^3} \frac{2(l^2+l+1)+(2l^2+1)\Lambda}{(l-1)(2l+1)} C^{l-1}_{m-\gamma}\,^1_{\gamma}\, ^l_m \left(\frac{\rho^{\prime}}{R}\right)^{l-2}  A^{\gamma \delta}_{l-2 m}(\Omega^{\prime}) P^{\prime}\medspace_{\gamma}\,^{\delta} \nonumber \\ 
c^f_{lm}&=&-\frac{1}{cR^3}\frac{(l+2)(2+\Lambda)}{l-1} \left (\frac{\rho^{\prime}}{R}\right)^{l-1}  C^l_{m-\gamma}\,^1_{\gamma}\,^l_m A^{\gamma \delta}_{l-1 m}(\Omega^{\prime}) P^{\prime}\thinspace_{\gamma}\thinspace^{ \delta}  
\label{free-sphere}
\end{eqnarray}
with
\begin{equation}
M(l)=2(l^2+1+l)+(2l^2+4l+3)\Lambda\ .
\end{equation}
For both boundary conditions the image displacements scale $\sim
1/R^2$ with the sphere radius and the VSH-expansion of $\vec u^b$
converges as $\sim l^2 (\rho \rho^{\prime})^l$.  Thus, higher
$l$-moments dominate if the dipole is close to the surface
$(\rho^{\prime}\rightarrow 1)$.  These are localized near the surface
and decay rapidly towards the sphere center.  We furthermore see, that
for a dipole close to the surface the convergence properties of the
series expansion are rather poor and more $l$-terms need to be
considered to approximate the displacement field near the surface.
Again clamped and free boundary induce opposing boundary fields as
indicated by the opposite signs of the expansion coefficients: a
clamped surface decreases $\vec u$ to zero at the boundary whereas a
free boundary enhances the displacements at the boundary.  In
\fig{fig:SphereInteract} we plot two examples for a deformed elastic
sphere with free boundaries under the action of a contraction dipole.

\begin{figure}
\begin{center}
\includegraphics[width=0.4\columnwidth]{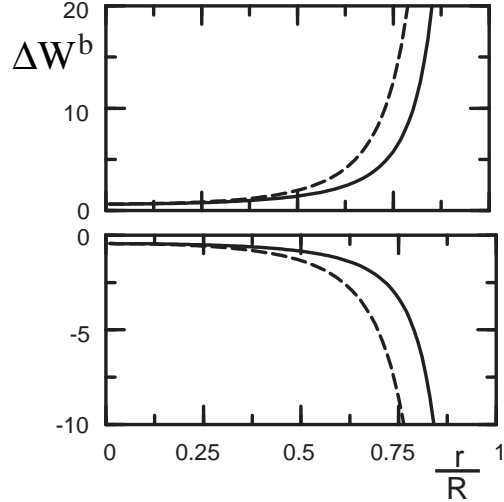}
\caption{\label{fig:SpherePNAS} Image interaction $\Delta W^b$ 
from \eq{interactionsphere} between the surface and a cellular force
dipole embedded in an elastic sphere of radius $R$ with $\nu=1/3$,
plotted in units of $P^2/ER^3$ as a function of distance $r/R$ to the
sphere surface and rescaled by 15/8. Curves above and below the
$r$-axis correspond to free and clamped boundary conditions,
respectively.  Solid and dashed line correspond to orientations
$\theta=\pi/2$ and $\theta =0$ with respect to the surface normal. As
for the halfspace, optimal cell orientation yields $\theta=0$
(clamped) and $\theta=\pi/2$ (free) respectively.}
\end{center}
\end{figure}

In order to calculate the change in effective stiffness sensed by a
contraction dipole at $\vec{r'}$ in an elastic sphere, we need
to contract the gradient-displacement tensor of the boundary induced
field with the dipole tensor. This is most conveniently done using the
spherical representation, i.e.:
\begin{equation}
\Delta W^b(\vec{r'}) = \frac{1}{2}P^{\alpha}\,_{\beta}u^b\,_{\alpha},^{\beta}(\vec{r}\rightarrow r^{\prime},\vec{r'})
=\frac{1}{2}P^{\alpha}\,_{\beta}(\bf e_{\beta}^{\dagger}\cdot\nabla)(e_{\alpha}\cdot \vec u^b)\ .
\label{trafo}
\end{equation}
Starting from the ansatz  Eq.~(\ref{ansatz}) for $\vec u$,
$u_{\alpha}\,^{\beta}(\vec{r},\vec{r'})$ can be derived by applying
the gradient formula for spherical harmonics \cite{b:edmo74}:
\begin{eqnarray}
\nabla \Phi(r) Y_{lm}(\Omega)&=&-\sqrt{\frac{l+1}{2l+1}}\left(\frac{d}{dr}-\frac{l}{r}\right)\Phi(r)\mathbf{Y_{ll+1m}}(\Omega)\nonumber \\&+&\sqrt{\frac{l}{2l+1}}\left(\frac{d}{dr}+\frac{l+1}{r}\right)\Phi(r)\mathbf {Y_{ll-1m}}(\Omega),
\end{eqnarray}
and furthermore the symmetry relationships of Clebsch Gordon 
coefficients \cite{b:edmo74}:
\begin{eqnarray}
C^{j_1}_{m_1}\,^{j_2}_{m_2}\,^{j_3}_{m_3}&=&(-1)^{j_2+m_2}\sqrt{\frac{2j_3+1}{2j_1+1}}C^{j_2}_{-m_2}\,^{j_3}_{m_3}\,^{j_1}_{m_1} \\
C^{j_1}_{m_1}\,^{j_2}_{m_2}\,^{j_3}_{m_3}&=&(-1)^{j_1+j_2-j_3} C^{j_1}_{-m_1}\,^{j_2}_{-m_2}\,^{j_3}_{-m_3} \nonumber.
\end{eqnarray}
We finally find:
\begin{eqnarray}
u^b\,_{\alpha}\,^{\beta}(\vec{r},\vec{r'})&=&\sum_{lm}R^l \frac{a_{lm}}{1+\Lambda} X^{\ast}\,^{\alpha \beta}_{lm}(\frac{r}{R},\Omega)  \\
&-&(2l+3)r^l A^{\ast}\,^{\alpha \beta}_{lm}(\Omega)\left(\frac{b_{l+2m}}{2} C^{l+1}_{m-\alpha}\,^1_{\alpha}\,^{l+2}_{m}+c_{l+1m} C^{l+1}_{m-\alpha}\,^1_{\alpha}\,^{l+1}_m\right). \nonumber
\label{interaction}
\end{eqnarray}
Note that the $m$-sums over $b_{lm}$ and $c_{lm}$ run in the intervals 
$[-l-2,l+2]$ and $[-l-1,l+1]$, respectively. The boundary induced change 
in stiffness sensed by a force dipole in an elastic sphere is then found 
by inserting the appropriate expansion coefficients $a_{lm}$, etc.
given in Eqs.~(\ref{clamped-sphere},\ref{free-sphere}) and
contracting $u_{\alpha}\,^{\beta}$ with $P^{\alpha}\,_{\beta}=P^{\prime}\,^{\alpha}\,_{\beta}$.
We may rewrite $\Delta W^b$ to indicate the important 
scaling laws of the interaction of the dipole with the sphere surface by:
\begin{equation} \label{interactionsphere}
\Delta W^b=\frac{P^2}{ER^3} f_{\nu}(\frac{r}{R},\theta)\ ,
\end{equation}
where $r$ is the distance to the sphere center and $\theta$ is the
dipole orientation with respect to the surface normal. The function
$f_{\nu}$ contains the sum over all angular momenta and does not vary
qualitatively as $\nu$ (or, equivalently, $\Lambda$) is varied.  With
regard to cell orientation, we find the same results as for the
elastic halfspace: cells will orient parallel (perpendicular) to a
free (clamped) surface, respectively. As shown in
\fig{fig:SpherePNAS}, we also find a similar result for the effect of
distance to the surface: for free (clamped) boundary conditions, a
small (large) distance to the sphere center is more favorable, since
the surface favors (disfavors) mechanical activity.  The new aspect
here is the role of the sphere radius $R$. Since $|\Delta W|$
increases when $R$ decreases, one can effectively rigidify (soften) a
material with a clamped (free) surface by reducing system size. For
the interaction of a physical dipole with the surface embedded in an
elastic sphere, we once more obtain the opposite results.  Dipoles are
attracted (repelled) and orient towards (away from) a free (clamped)
surface.

\begin{figure}
\begin{center}
\includegraphics[width=\columnwidth]{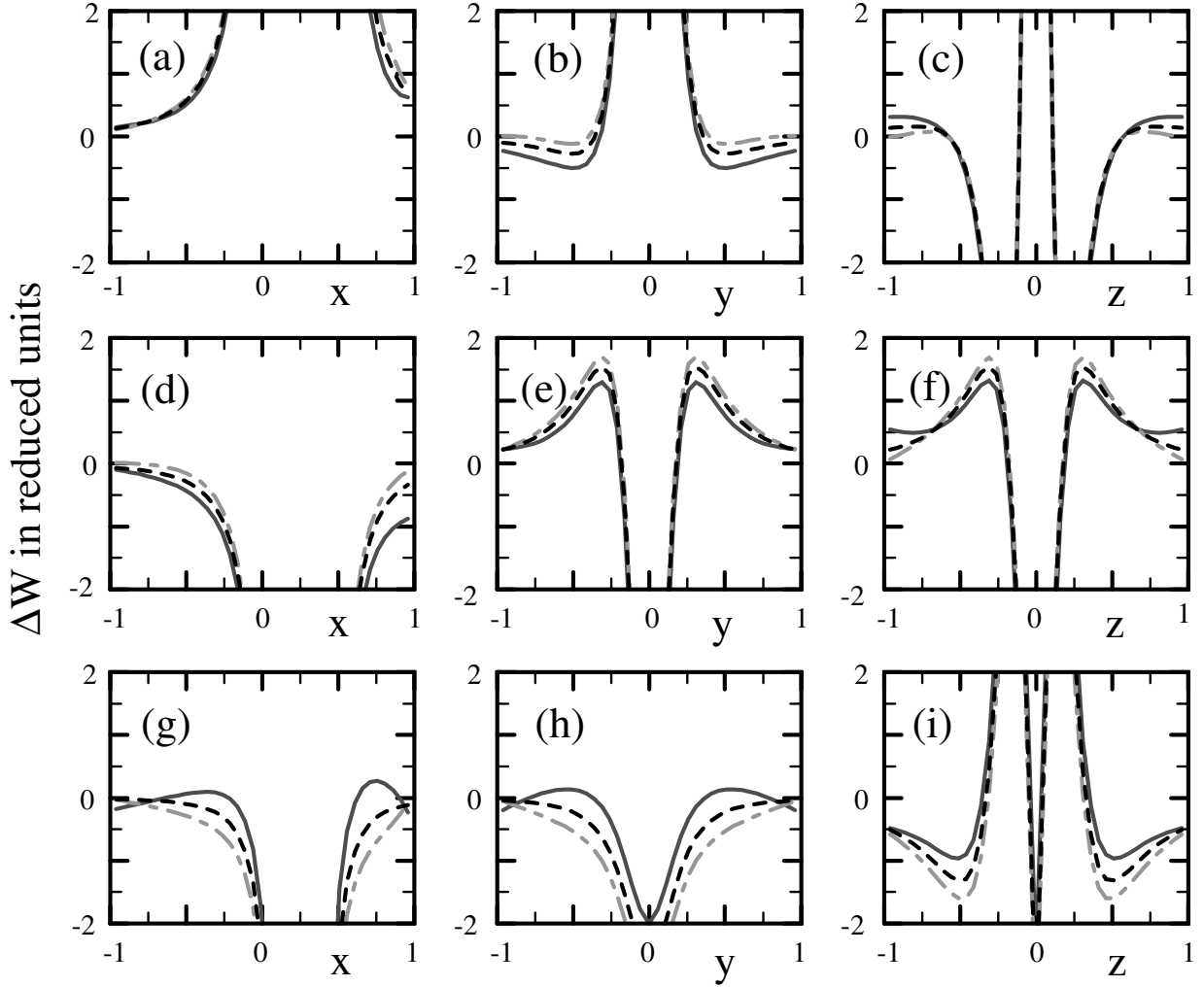}
\caption{\label{fig:DipoleInteract} Cellular dipole-dipole interactions
$\Delta W=\Delta W^{\infty}+\Delta W^{b}$ in an elastic sphere
$(\Lambda=2)$ in units of $PP^{\prime}/cR^3$ for clamped (dashed gray)
and free (full gray) boundary conditions.  A z-dipole is fixed at
$\vec{r}=(\frac{R}{4},0,0)$. A x-dipole (a,b,c), y-dipole (d,e,f) and
z-dipole (g,h,i) is moved along the coordinate axes.  The boundary
condition introduces corrections to the interaction in infinite medium
(black line) that vary on the macroscopic scale.  The boundary term
dominates close to the surface and in some cases introduces new maxima
or minima in the interaction landscape.}
\end{center}
\end{figure}

So far we have considered the interaction of a force dipole with the
boundary. One may extend our model of cell-cell interactions to cells
embedded in finite geometries and study how their boundaries alter the
interaction between cells.  In an elastic sphere containing many
cells, we can separate the contributions to the effective stiffness
into a contribution from the boundary induced field, i.e.\ a
cell-surface interaction as discussed above, and a contribution from
the elastic fields of other cells embedded in the sphere, i.e.\ a
cell-cell interaction term. This contribution is modified with respect
to the interaction term in infinite medium, Eq.~(\ref{3DW}), by a
boundary mediated interaction term. The indirect interaction term is
given by contracting the dipole tensor of the first dipole with the
image strain caused by the second dipole.  The most important result
here is that the image term varies on the macroscopic scale $R$.  For
physical dipoles, elastic interactions in finite sized geometries have
been studied extensively, in particular for \textit{isotropic}
dipoles, that do not interact in infinite medium and where the
interaction between dipoles is mediated solely via the boundary
\cite{e:wagn74}.  By setting $P_{\alpha}\,^{\beta}=\delta_{\alpha
\beta}$ our results specialize to the interaction of isotropic dipoles
in an elastic sphere
\begin{equation}
\Delta W^b(\vec{r}, \vec{r^{\prime}})=
-V^b(\vec{r},\vec{r'})=
\sum M_{l} r^l r^{\prime}\,^l Y^{\ast}_{lm}(\Omega) Y_{lm}(\Omega^{\prime})
\end{equation}
with
\begin{eqnarray}
M_l^{\rm f}&=&\frac{PP^{\prime}}{cR^3} \frac{2 (2l+3) (l+1)(l+2)}{2(l^2+l+1)+\Lambda (2l^2+4l+3)}\\
M_l^{\rm c}&=&-\frac{PP^{\prime}}{cR^3}\frac{2l+3}{(l+1)(1+(\Lambda+3)l)},
\end{eqnarray}
for free and clamped boundaries, respectively. These results can be
shown to be identical with the ones for isotropic dipoles previously
reported in Ref.~\cite{e:wagn74}.  Note that the interaction of
physical isotropic defects is always attractive (repulsive) for
isotropic dipoles in a free (clamped) sphere.  Due to the macroscopic
interaction range of isotropic physical dipoles the indirect
interactions lead to structure formation on the macroscopic scale
(\textit{macroscopic modes}), e.g.\ in hydrogen-metal alloys
\cite{e:wagn74}. For anisotropic dipoles the image interaction 
introduces corrections to the direct interaction term, which vary on
the macroscopic scale. In \fig{fig:DipoleInteract} we plot the
interaction of two anisotropic dipoles in infinite medium and the
modified interactions in clamped and free spheres, respectively.  For
example, the image correction in a free sphere for two parallel
z-dipoles (one placed at the sphere center) along the x-axis reads
\begin{equation}
\Delta W^b(x) = \frac{PP^{\prime} \left[ (112 + 352 \Lambda + 370 \Lambda^2 + 135 \Lambda^3) 
- 12 (7 + 4 \Lambda) (2 + 5 \Lambda + 3 \Lambda^3) \left(\frac{x}{R}\right)^2\right]}
{4 (2 + 3 \Lambda) (14 + 19 \Lambda) \pi c R^3}\ .
\end{equation}
For $\Lambda \rightarrow \infty$ ($\nu = 1/2$), this becomes
\begin{equation}
\Delta W^b(x) = \frac{PP^{\prime}}{76 \pi \mu R^3}\left[45-48\left(\frac{x}{R}\right)^2 \right]\ .
\end{equation}
For a clamped sphere, we find
\begin{equation}
\Delta W^b(x) = \frac{PP^{\prime} \left[ - (686 + 280 \Lambda + 24 \Lambda^2) 
+ 45 (1 + \Lambda) (7 + 4 \Lambda) \left(\frac{x}{R}\right)^2\right]}
{120 (7 + 2 \Lambda) \pi c R^3}
\end{equation}
which for $\nu = 1/2$ becomes
\begin{equation}
\Delta W^b_{\rm c}(x) = \frac{PP^{\prime}}{20 \pi \mu R^3}\left 
[-2+15\left(\frac{x}{R}\right)^2\right]\ .
\end{equation}
Again we find that clamped and free surface result in opposite
effects.  On the microscopic scale (i.e. for small cell-cell
distances), the direct interaction dominates. For macroscopic cell
separations, the boundary term introduces significant contributions
that dominate over the direct term close to the surface.  For some
cases, the boundary can induce new maxima or minima in the
dipole-dipole interaction landscape.  Note that for a full treatment,
the dipole-surface interactions have to be included.  In conclusion,
in contrast to isotropic dipoles, structure formation of anisotropic
dipoles is dominated by effects on cellular and elastic scales, which
result from direct interactions.  Since they compete with boundary
induced effects on a macroscopic scale, in general we expect
hierarchical structures.

\subsection{Summary example section}

In the second part of this paper, we applied the general formalism
from the first part to different situations of interest. In general,
we found that physical and cellular force dipoles interact in opposite
ways with each other, external strain field or sample boundaries,
because $V_t = - W$. For example, physical anisotropic force dipoles
on top of thick elastic films or in infinite elastic material locally
prefer the T-configuration (for Poisson ratio $\nu = 1/2$), while
cellular anisotropic force dipoles align in strings (independent of
the value for $\nu$). The predicted structure formation for physical
force dipoles and active cells is similar to the ones of electric
quadrupoles and dipoles, respectively. We also found that in general,
free and clamped boundaries will have opposite effects.  For example,
cellular anisotropic force dipoles are repeled and attracted by free
and clamped boundaries, respectively. In the vicinity of these
boundaries, they will align in parallel and perpendicular,
respectively. In general, all the interaction laws derived here show
the universal scaling $W \sim (P^2 / E l^3) f_{\nu}(\theta_i)$, where
$f$ is a non-trivial function of Poisson ratio $\nu$ and the different
angles $\theta_i$, which has to be calculated for each situation of
interest.  Except for the case of external strain, the cellular force
pattern interacts with itself (case of boundaries) or with another
cellular force pattern (case of elastic interaction of cells),
therefore $W \sim P^2$. The scaling $W \sim 1/ l^3$ is typically for
force dipoles. Here the length $l$ can either be distance (e.g.\
between cell and boundary or between two cells) or sample size (in the
elastic sphere). Finally, $W \sim 1/E$. Although $W$ decreases with
increasing Young modulus $E$, that is elastic effects become smaller,
at the same time mechanical activity of cells usually increases. For
this reason, we expect that there exists a range of optimal values for
$E$ for which the elastic effects in cell adhesion described here
should be most pronounced (possibly around $E = kPa$, the
physiological order of magnitude for cell and tissue stiffness).

Although our modeling focuses on the role of strain and stress in the
extracellular environment, we also need a model for the typical force
pattern of mechanically active cells. Since the minimal system for
contractile activity of adherent cells is one stress fibers connecting
two focal adhesions, we introduced the concept of force dipoles into
the physics of cells \cite{uss:schw02a}. From a technical point of
view, this allowed to make contact to a large body of results on
physical force dipoles in deformable media.  Our theory reproduces
known results for physical force dipoles, in particular the elastic
image interaction of isotropic force dipoles in an free elastic sphere
\cite{e:wagn74}. The corresponding calculation for anisotropic dipoles
has been done before by Hirsekorn and Siems \cite{e:hirs81}, but only
for the free surface.  Here we extended this calculation to the
clamped case. Moreover, in order to predict single cell effects, we
also calculated the interaction between dipole and surface for both
types of boundary condition. In contrast to the elastic sphere, for
the elastic halfspace an image system for the effect of force
monopoles is known \cite{e:mind36}.  Here we used the solution given
by Walpole \cite{e:walp96} and adapted it for the case of force
dipoles.

As reported earlier, our predictions for cell organization in soft
media are in excellent agreement with experimental observations
\cite{uss:bisc03a}. Our theory not only contributes to a better
understanding of physiological processes involving mechanical activity
of cells (including tissue maintenance, wound healing, angiogenesis,
development and metastasis), in the future it also might be used to
predict cell behaviour in artificial tissues, close to implants and on
compliant biosensors. Moreover, the orientation response of regulated
cells as described here might be used to distinguish between healthy
and diseased conditions. It is important to note that the main success
of our model results from the fact that we focus on the role of stress
and strain in the environment, which allows to use the concepts of
linear elasticity theory and to make minimal assumptions about
cellular regulation.  In the future, our theory might be complemented
by models for cell morphology and the dynamics of focal
adhesions. Moreover, until now we did not address in detail the issue
of structure formation within large communities of cells, although
this might be of large importance for development, when large groups
of mechanical active cells are known to move in concert.

\textbf{Acknowledgments:} I.B.B. thanks Thomas Pfeifer for helpful 
discussions. S.A.S. acknowledges support by the Schmidt Minerva
Center, the Israel Science Foundation and the U.S. Israel Binational
Science Foundation. I.B.B. and U.S.S. are supported by the German
Science Foundation through the Emmy Noether Program.


\end{document}